\documentclass[useAMS,usenatbib]{mn2e}

\usepackage{graphicx,subfigure,natbib}

\def\reff@jnl#1{{\rm#1\/}}
\def\apj{\reff@jnl{ApJ}}       
\def\mnras{\reff@jnl{MNRAS}}   
\def\nat{\reff@jnl{Nature}}    
\def\aap{\reff@jnl{A\&A}}      

\newcommand{\that}{\hat{t}}
\newcommand{\rhat}{\hat{r}}
\newcommand{\thetahat}{\hat{\theta}}
\newcommand{\phihat}{\hat{\phi}}

\newcommand{\sDelta}{\sqrt{\Delta}}
\newcommand{\ddt}{\partial_t}
\newcommand{\ddr}{\partial_r}
\newcommand{\ddtheta}{\partial_\theta}
\newcommand{\ddphi}{\partial_\phi}
\newcommand{\tdot}{\dot{t}}
\newcommand{\rdot}{\dot{r}}
\newcommand{\thetadot}{\dot{\theta}}
\newcommand{\phidot}{\dot{\phi}}
\newcommand{\Phidot}{\dot{\Phi}}
\newcommand{\alphdot}{\dot{\alpha}}
\newcommand{\betadot}{\dot{\beta}}
\newcommand{\prh}{\hbox{\bf \em h}}
\newcommand{\prD}{\hbox{\bf \em D}}
\newcommand{\prob}{\hbox{Pr}}
\newcommand{\clr}{\mathcal{R}}
\newcommand{\boldx}{\hbox{\bf \em x}}
\newcommand{\boldz}{\hbox{\bf \em z}}
\newcommand{\nn}{\nonumber}

\title[Inferring the coronal flaring patterns in AGN]{Inferring the
  coronal flaring patterns in AGN from reverberation maps}
\author[R. Goyder and A. N. Lasenby]{R. Goyder\thanks{E-mail: rg200@mrao.cam.ac.uk} and A. N. Lasenby\\
  Astrophysics Group, Cavendish Laboratory, Madingley Road, CB3 0HE}
\begin{document}
\pagerange{\pageref{firstpage}--\pageref{lastpage}} \pubyear{2002}
\maketitle
\label{firstpage}
\begin{abstract}
  The relativistically broadened iron K-$\alpha$ line at 6.4 keV
  observed in the Seyfert 1 galaxy MCG--6-30-15 has provided a probe
  of the strong-gravity environment near a black hole, in particular
  suggesting that it is rapidly spinning. An important variable in
  such analyses is the geometry of the illuminating source. We present
  a new technique which constrains this geometry based on the spectral
  line shape, based on a model of discrete, point-like flaring regions
  in the X-ray corona. We apply it to simulated reverberation maps and
  give examples of successful reconstructions of complex coronal
  flaring patterns. For time-averaged spectral lines the problem is
  highly degenerate, and so its inversion more challenging. We
  quantify this degeneracy and give a measure of the spatial accuracy
  of the method in this case, before checking that it is consistent
  with the existing picture of MCG--6-30-15 by applying it to recent data from
  {\em XMM-Newton}.
\end{abstract}
\section{Introduction}\label{sec:introduction}
The broad, skewed emission line around 6.4 keV observed in the Seyfert
1 galaxy MCG--6-30-15 was first resolved with ASCA
\citep{1995Natur.375..659T}. A widely accepted model for the
production of this line is one where hard X-rays are reprocessed by
the innermost regions of a cold, thin accretion disc around a
super-massive black hole
\citep{1991MNRAS.249..352G,1991A&A...247...25M}. X-ray photons emitted
from this material suffer large gravitational redshifts, special
relativistic beaming and Doppler shifts, combining to give a
characteristic shape which is in good agreement with the data. In
addition, a number of alternative models were considered by
\cite{1995MNRAS.277L..11F}, but a convincing rival candidate was not
found.

The extended red wing of this line
\citep{2001MNRAS.328L..27W,2002MNRAS.335L...1F,fabianvaughan2003} has
provided the first direct observational probes of the strong gravity
environment near a black hole and has been used to constrain
parameters such as the inclination of the accretion disc and the black
hole's spin \citep{1996MNRAS.282.1038I,1997MNRAS.288L..11D}. The
geometry of the illuminating source of hard X-rays, or corona, is a
critical variable in such calculations, as illustrated by the work of
\cite{1997MNRAS.288L..11D}, where the authors assumed that the X-ray
source was planar, hugging a disc which extended down to the innermost
stable circular orbit in the Kerr metric. Within this framework, they
showed that the line profile obtained during a 4 day ASCA observation
of MCG--6-30-15 in 1994 constrained the spin parameter of the black
hole to be $>0.94$ with about $90\%$ confidence.

The central regions of an accretion disc are, however, likely to be
subject to a large number of magnetohydrodynamical instabilities with
the result that a planar geometry for the hard X-ray source is a poor
approximation. An alternative approach was taken by
\cite{1997ApJ...488..109R}. They were able to fit the same data by
assuming a Schwarzschild black hole where the X-ray source was a
point-like flare located a few gravitational radii above the hole on
the symmetry axis of the disc.  The bulk of the emitted radiation
would then come from material on plunge trajectories within the
innermost stable orbit radius.  Shortly afterwards,
\cite{1998MNRAS.300L..11Y} showed that a geometry of this type
predicts a large absorption edge in the spectrum, due to the line
emitting matter being highly ionised.  As no such edge has been
observed, the data seem to favour the Kerr model, although
tentatively.

Of particular interest for the future are time-dependent scenarios,
where a flash of radiation sweeps over the disc producing line
emission in different locations and causes the line shape to change
with time \citep{1999ApJ...514..164R}. Future X-ray satellites such as
{\em Constellation-X} offer the possibility of observing these
time-dependent spectra, or reverberation maps \citep{2000ApJ...529..101Y}.  However, analyses of
such data depend on the coronal flaring pattern, and most previous
work has concentrated on predicting the spectra which result from
one or two dominant flaring events in the corona. A more likely scenario, however,
is one where the line profiles in reverberation maps result from an
ensemble of overlapping flares, distributed in time as well as space,
and so a treatment of more complicated coronal configurations such as
these is vital. Here we aim to develop a technique for acquiring
knowledge of this pattern from spectral data, constraining the
distribution of flares in the corona based on the shape of a spectral
line. We consider an ensemble of coronal flares of arbitrary number,
instead of assuming that a single flare dominates the emission.

In general, models are commonly fitted to data via the $\chi^2$
statistic, the best fit parameters corresponding to the minimum of a
Gaussian distribution. With noisy data this can result in over-fitting
where many different parameter values produce a similarly good fit to
the data. In this case some form of regularisation is often applied,
such as the Maximum Entropy Method \citep{1978Natur.272..686G}. This
amounts to making some extra basic assumptions about the object one is
trying to infer, in order to pick out a preferred model from a group
which are not distinguished by the data alone. The exact physics of a
magnetically flaring corona is complicated, and so rather than trying
to treat it directly, we can make some progress by making some very general,
minimal assumptions about its nature.

Given the idea that individual X-ray flares provide the illumination
for the production of the iron line, we attempt to infer the
parameters of a set of discrete, point-like sources, which are equally
likely to be found at any point over the disc. We assume that the
number of these sources is Poisson distributed, so that if the data
can be satisfied with a lower number of flares then it will.  Finally,
we assume an exponentially decaying distribution for the strengths of
these sources, so that stronger flares are not favoured in the
reconstruction. This set of assumptions corresponds to a technique
known as `Massive Inference' \citep{massinf} and is quite general. It
can be applied to any situation which can be modelled as arising from
a number of point source like objects, however abstract their physical
interpretation might be.

In this paper we begin by introducing the Massive Inference technique
and its numerical implementation, in
Sections~\ref{sec:mass-infer-prior} and
\ref{sec:computational-method}. In Section~\ref{sec:line-prof-calc} we
then discuss the physical details of our model for the forward
problem, that of predicting the iron line shape as a function of the
coronal geometry and other system parameters. After describing the
calculation with which we compute the line profile due to a single
flare we proceed to discuss how these lines can be combined to yield a
transfer function which maps the parameters of interest to a composite
line profile, in Section~\ref{sec:transfer-functions}. We then show
some examples of the technique's application to simulated data: to
reverberation maps in Section~\ref{sec:time-depend-invers} and
time-averaged spectral lines in Section~\ref{sec:time-indep-invers}.
The transfer function for the latter case is analysed in
Section~\ref{sec:linear-dependence} before treating data from an {\em
  XMM-Newton} observation of MCG--6-30-15 in
Section~\ref{sec:application-xmm-data}. Finally, in
Section~\ref{sec:conclusion} we give our conclusions and discuss the
application of our technique to possible future data.

\section{Massive inference prior}\label{sec:mass-infer-prior}
It is standard practice in scientific data analysis to fit the
parameters of a model to a dataset using a $\chi^2$ minimisation.
Often, especially at high levels of noise, a unique minimum is not
clearly preferred by the data and there is ambiguity between different
sets of parameters. We can still progress in a such a problem
though, by making extra assumptions about the parameters, which are
believed to hold independently of the data. These are encoded in a
`prior' probability distribution, which modifies the previous Gaussian
distribution to one where a clearly defined minimum exists, and so a
`best' answer can be found, although it is of course still dependent
upon the validity of the extra assumptions.

A well known example of this idea is the `Maximum Entropy Method'
\citep{gull89}, which has been used extensively in many areas of
physics and astronomy, most notably in image processing. Here, the
parameters are the fluxes collected in a set of pixels. By assuming
that the fluxes are all positive and additive \footnote{Fluxes are
  necessarily positive, and `additive' means that the flux received
  across two non--overlapping patches is the sum of the individual
  fluxes.}, one arrives at the conclusion that the image, in the
absence of any data, has a prior probability distribution given by
\begin{equation}\label{eq:72}
  \prob(\prh) = \exp(-\alpha_r S)
\end{equation}
where $S$ is the entropy defined as
\begin{equation}
S({\prh}) = \sum_{i}^{N} \left(h_i - m_i - h_i\log\left(\frac{h_i}{m_i}\right)\right)
\end{equation}
and where $N$ is the number of pixels in the image, $h_i$ (the
$i^{th}$ element of the vector $\prh$) is the flux in the $i$th pixel
and the $\{m_i\}$ represent our `best guess' at the answer in the
absence of any data -- often a uniform distribution.
$\alpha_r$ represents the weight we wish to give the Maximum Entropy
assumption relative to the data. Maximising the entropy essentially
corresponds to imposing the least possible prejudice on the image, and
the best image is the one which strikes the optimum compromise between
satisfying the data, by minimising the $\chi^2$ statistic, and
maximising the entropy.

In this work, instead of using a maximum entropy prior, we instead
follow the `Massive Inference' route, which can be expected to be
particularly appropriate to the case of a small number of point-like
sources. To provide a rigorous framework in which we can combine our
prior assumptions with line profile data, we use a precise statement
of conditional probability -- Bayes Theorem.  According to Bayes
Theorem, a hypothesis $\prh$ given some data $\prD$ has a {\em
  posterior} probability distribution $\prob(\prh | \prD)$ given by
the product of the {\em likelihood} $\prob(\prD | \prh)$ and the {\em
  prior} $\prob(\prh)$ normalised by the {\em evidence} $\prob(\prD)$
\begin{equation} \label{bayes}
\prob(\prh | \prD) = \frac { \prob(\prh) \; \prob(\prD | \prh) } { \prob(\prD) }.
\end{equation}
We assume Gaussian random errors, so that the likelihood takes the
form
\begin{eqnarray}
\prob(\prD | \prh) & \propto & e^{-\chi^2/2} \\
\chi^2 & = & \sum_i^M[\prD-\clr(\prh)]_{i} \, \hbox{\bf \em H}_{ii} \, [\prD-\clr(\prh)]_i,
\end{eqnarray}
where $M$ is the number of data points in the vector $\prD$ and the
covariance matrix $\hbox{\bf \em H}$ is diagonal because the errors
are independent. In this work, the hypothesis $\prh$ is comprised of
the number, strengths and locations of a group of X-ray flares, and
the data $\prD$ are the iron line X-ray fluxes received in each energy
bin around $6.4$ keV. $\clr$ is the transfer function which provides a
map from the parameter space to the data, so that $\clr(\prh)$ are the
(noiseless) `mock data' corresponding to a set of parameters $\prh$.
The prior probability $\prob[\prh(N,\boldx,\boldz)]$ consists of the
product of three separate priors for each of $N$ (the number of
flares), $\boldx$ (an $N$ element vector of flare positions) and
$\boldz$ (an $N$ element vector of flare strengths). $N$ is assigned
either a Poisson distribution
\begin{equation} \label{poisson}
\prob(N) = \frac{e^{-\alpha}\alpha^N}{N!} \qquad
\langle N\rangle = \alpha\pm\sqrt{\alpha}
\end{equation}
with mean $\alpha$, or a wider, geometric distribution
\begin{eqnarray} \label{geometric}
\prob(N) &=& (1-r)r^N, \quad r=\frac{\alpha}{1+\alpha} \\
\langle N \rangle &=& \alpha \pm \sqrt{\alpha(1+\alpha)}
\end{eqnarray}
depending on how accurately the mean number of flares $\alpha$ can be
estimated \citep{miman}. In practice we always use the latter, with an
additional, exponentially decaying prior on $\alpha$ itself. The flare
positions are assigned a prior which is constant over the allowed
locations (in practice covering the central regions of the accretion
disk) and zero otherwise. Their strengths an exponentially decaying
distribution
\begin{equation}
\prob({\boldz}) = \prod_{i}^{N} \frac{e^{-z_i/q_i}}{q_i}
\end{equation}
where the decay constants $q_i$ are chosen internally based on the
overall scale of the data, and so do not feature in the analysis. This
form avoids extremely strong flares being favoured in the posterior.

The full posterior probability distribution can then be written
\begin{eqnarray}
\prob(\prh | \prD) & \propto & \frac{e^{-\alpha}\alpha^N}{N!} \; 
\prod_{i}^{N} \frac{e^{-z_i/q_i}}{q_i} \;
e^{-\chi^2/2} \\
\chi^2 & = & \sum_i^M[\prD-\clr(\prh)]_{i} \, \hbox{\bf \em H}_{ii} \, [\prD-\clr(\prh)]_i
\end{eqnarray}

Finding the value of $\prh$ which maximises this distribution will
give the `best' set of parameters for a given dataset. This, however,
is only a fraction of the total amount of information contained in the
answer. In this work we gain a knowledge of the full posterior
distribution by sampling with a simulated annealing algorithm
\citep{miman} described in more detail in
Section~\ref{sec:computational-method}. We can then calculate any
moment of the distribution, such as the mean and standard deviation,
rather than just finding the best-fitting model.

As mentioned above, this form of posterior is quite general -- in this
paper, the `locations' of the sources are the spacetime coordinates of
coronal flaring events, but this need not be the case. The physical
interpretation of the parameter space is entirely determined by the
form of the transfer function $\clr$ which provides a map from $\prh$
to the data (without noise)
\begin{equation} \label{h2D}
\clr: \prh \mapsto \prD.
\end{equation}
In Section~\ref{sec:line-prof-calc} we consider the detailed physics
which is included in this object, although the end result of the
calculation is simply a linear mapping from the three objects
contained in $\prh$ (ie $N$, $\boldx$ and $\boldz$) to the data
$\prD$.
\section{Line profile calculation}\label{sec:line-prof-calc}
The `forward problem' which we attempt to invert in this paper is that
of predicting the shape of the iron K$\alpha$ spectral line at 6.4
keV. We perform explicit ray-tracing in Kerr gravity which accounts
for all relativistic beaming and Doppler shifting effects, using the
geodesic equations given in the appendix.

We adopt a `lamp-post' geometry, where a point source orbits above an
optically thick, geometrically thin accretion disc, consisting of
particles on circular, time-like prograde geodesics in the equatorial
plane.  Its coordinates are $t_s, \rho_s, \phi_s, h_s$ in a
cylindrical polar system, where $t_s$ and $\phi_s$ are Boyer-Lindquist
coordinates and $\rho_s$ and $h_s$ are related to the Boyer-Lindquist
coordinates via $r_s^2 = \rho_s^2 + h_s^2$, $\tan\theta_s =
h_s/\rho_s$. Except where explicitly stated otherwise, the `radius' of
the source refers to its value of $\rho_s$, the distance from the
rotation axis of the black hole. As the source is located above the
plane, it is not in free-fall, rather, it remains above the same point
on the disc throughout its motion. The precise details of its
trajectory are also given in the appendix. The disc material orbits
with a period given by
\begin{equation}
  \label{eq:orbit_times}
  T = \frac{a}{c^2}+\sqrt{\frac{r^3}{GM}} = \left(\tilde{a}+\tilde{r}^{3/2}\right)\frac{GM}{c^3}
\end{equation}
where, in the first part of this equation we use conventional (S.I.)
units (where $GM/rc^2$ is dimensionless), and in the second part,
$\tilde{r}$ is measured in units of $r_g = GM/c^2$ and $\tilde{a}$ in
units of $GM/c$. The disc's inner radius is chosen to be that of the
marginally stable orbit and the disc extends outward to $100\,r_g$. We
set the black hole's angular momentum parameter $a$ to 0.998 so that
the accretion disc allows emission within $6\,r_g$
\citep{1995Natur.375..659T,2001MNRAS.328L..27W,2002MNRAS.335L...1F,fabianvaughan2003}.
In all the plots in this paper, the disc rotates in an anti-clockwise
sense.

We assume the source emits isotropically in its rest frame a power-law
spectrum with a specific luminosity $ \propto\nu^{-\alpha}$ erg
s$^{-1}$ Hz$^{-1}$. In practice we enforce this by choosing the
initial directions of the emitted photons such that they each pass
through the centre of one pixel of a HEALPix\footnote{Hierarchical
  Equal Area isoLatitude Pixellisation, see
  http://www.eso.org/science/healpix.} sphere. We can then compute the
incident flux at any point on the disc, as measured in the rest frame
of the disc matter \citep{2001ApJ...561..660L,2000MNRAS.312..817M}, to
be
\begin{equation} \label{eq:1}
  F_{in} = \frac{\nu_d^{-\alpha}}{(1+z_{sd})^{1+\alpha}}
\end{equation}
where $z_{sd}$ is the redshift suffered by a photon arriving at that
point and $\nu_d$ is the frequency as measured by an observer
co-moving with the disc. To find the flux that leaves the disc, we use
the following semi-analytical approximation for the effect of the
photons' incident energy and angle on the number emitted as iron
K$\alpha$ line radiation, which are derived from the Monte Carlo
simulations of the detailed atomic physics occurring in a cold,
non-ionised disc \citep{1991MNRAS.249..352G}:
\begin{equation}
  \label{eq:2}
  N_{out} = g(\theta_i) \int_{\nu_t}^{\nu_m} N_{in}(\nu)
  f(\nu) d\nu
\end{equation}
where
\begin{eqnarray}
\label{fg1}
f(\nu) & = & 7.4 \times 10^{-2} + 2.5\,\exp\left(
-\frac{\nu-1.8}{5.7}\right) \\
\label{fg2}
g(\theta_i) & = & 6.5 - 5.6\cos\theta_i + 2.2\cos^2\theta_i
\end{eqnarray}
and where $\nu$ is measured in keV and $\nu_t$ is the threshold for
triggering fluorescent emission. Above $\nu_m = 30$ keV this
approximation breaks down, but there is a negligible contribution from
this region to the total flux.  We obtain the following expression for
the flux of iron K$\alpha$ line photons leaving the disc
\begin{equation}\label{eq:4}
  F_{out} \propto \delta_{\nu_\alpha}\,\nu_d\,g(\theta_i)\frac{1}{(1+z_{sd})^{1+\alpha}}\int_{\nu_t}^{\nu_m}f(\nu_d)\nu_d^{-(1+\alpha)}d\nu_d
\end{equation}
We assume that the iron line radiation leaves the disc isotropically
and is collected in a grid of $2000^2$ equal area pixels (modelling
those of a CCD detector) in an asymptotically flat region of spacetime
(in practice $1000\,r_g$), at a location inclined at $30^\circ$ to the
disc, in keeping with appropriate parameters for MCG--6-30-15. To
calculate the change in intensity $I_\nu$ of radiation on its journey
from the disc to the observer we use the Liouville invariant
(e.g. Misner et al. 1973)
\begin{equation} \label{liou}
\frac{I_\nu}{\nu^3} = \mathrm{constant},
\end{equation}
where the change in frequency is computed for each pixel in the
detector by integrating photon trajectories backwards in time until
they hit the disc or are lost into the black hole
\citep{1991ApJ...376...90L,1997MNRAS.288L..11D}. The power collected
in one pixel in the finite frequency range $\Delta\nu$ is then
\begin{equation}\label{eq:5}
  W_{\Delta\nu} \propto \frac{1}{(1+z_{do})^4} \,F_{out}(r,\phi)
\end{equation}
where we have made the position dependence of the emissivity over the
disc explicit and assumed isotropic emission from the disc. We
integrate over the entire solid angle subtended by the disc at the
observer by summing the contribution from each pixel. The contribution
to the total flux of each photon is then binned in both energy and
arrival time, to build up a time-dependent spectrum.
\subsection{Transfer functions}\label{sec:transfer-functions}
\begin{figure*}
  \centering
  \subfigure[]{
    \begin{minipage}{8cm}
      \centering
      \label{fig:timedepgf}
      \includegraphics[height=8cm,angle=-90]{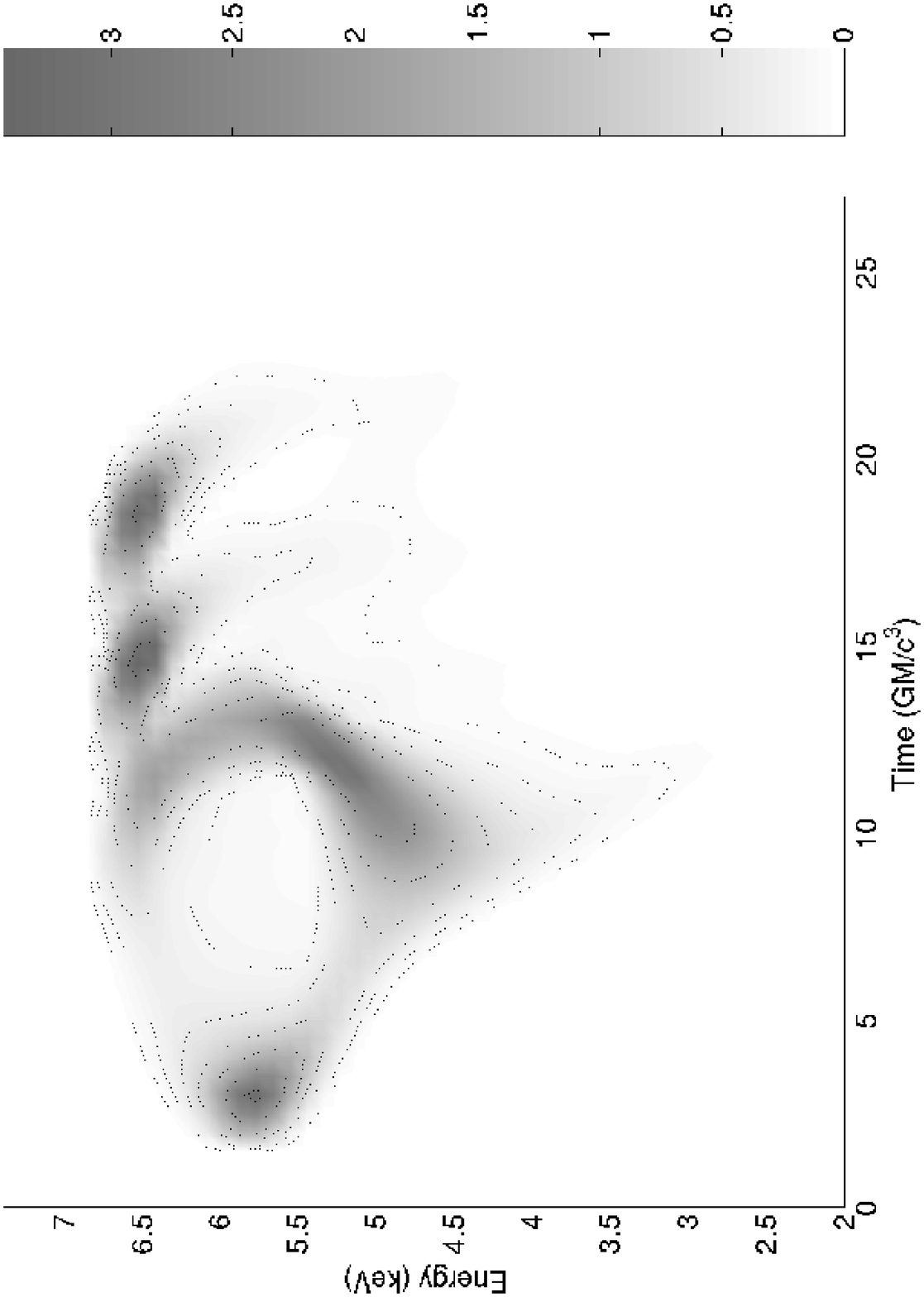}
    \end{minipage}
    }
  \hspace{1cm}
  \subfigure[]{
    \begin{minipage}{8cm}
      \centering
      \label{fig:ringgf}
      \includegraphics[height=8cm,angle=-90]{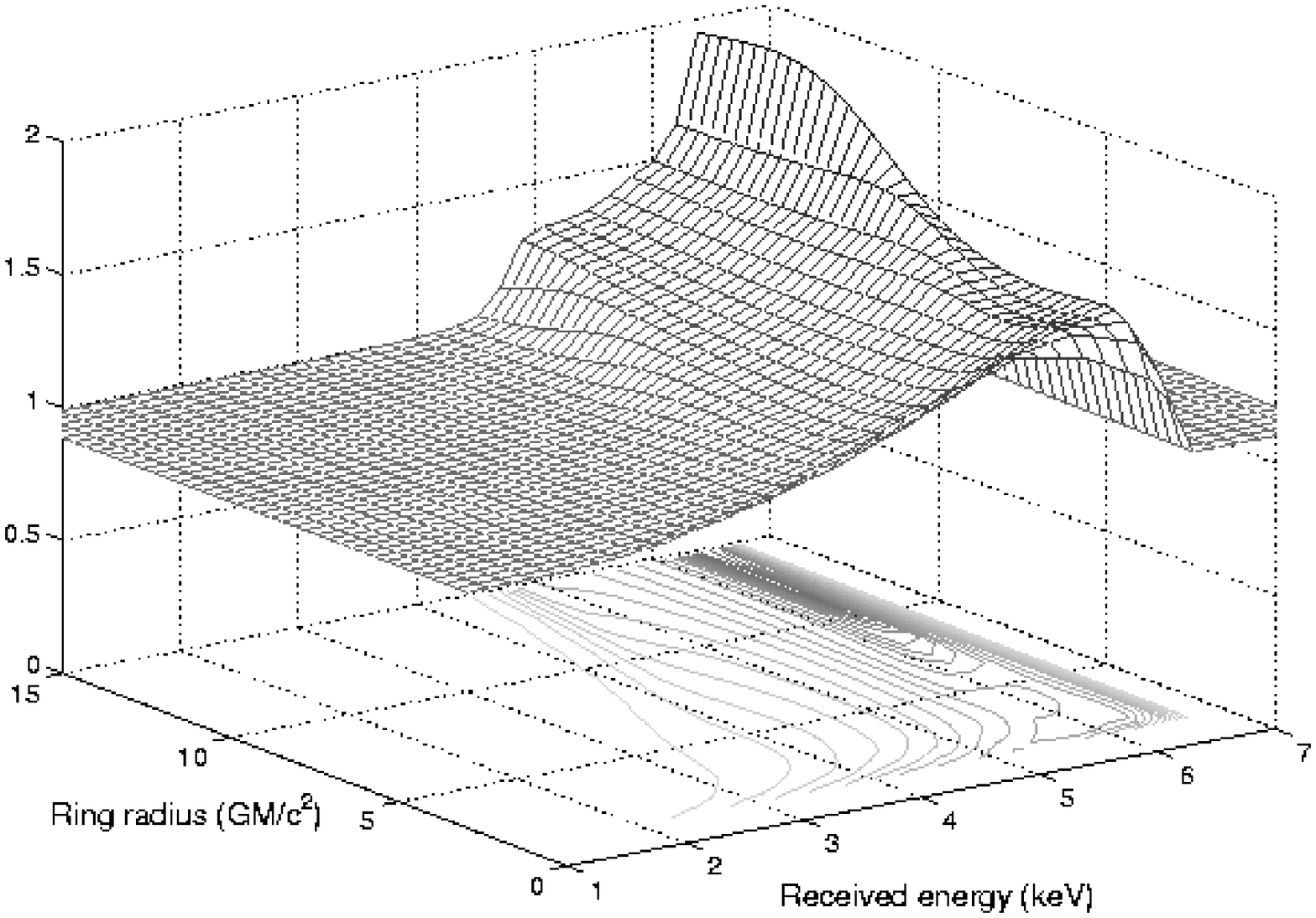}
    \end{minipage}
    }
  \subfigure[]{
    \begin{minipage}{8cm}
      \centering
      \label{fig:pointgfrad1}
      \includegraphics[height=8cm,angle=-90]{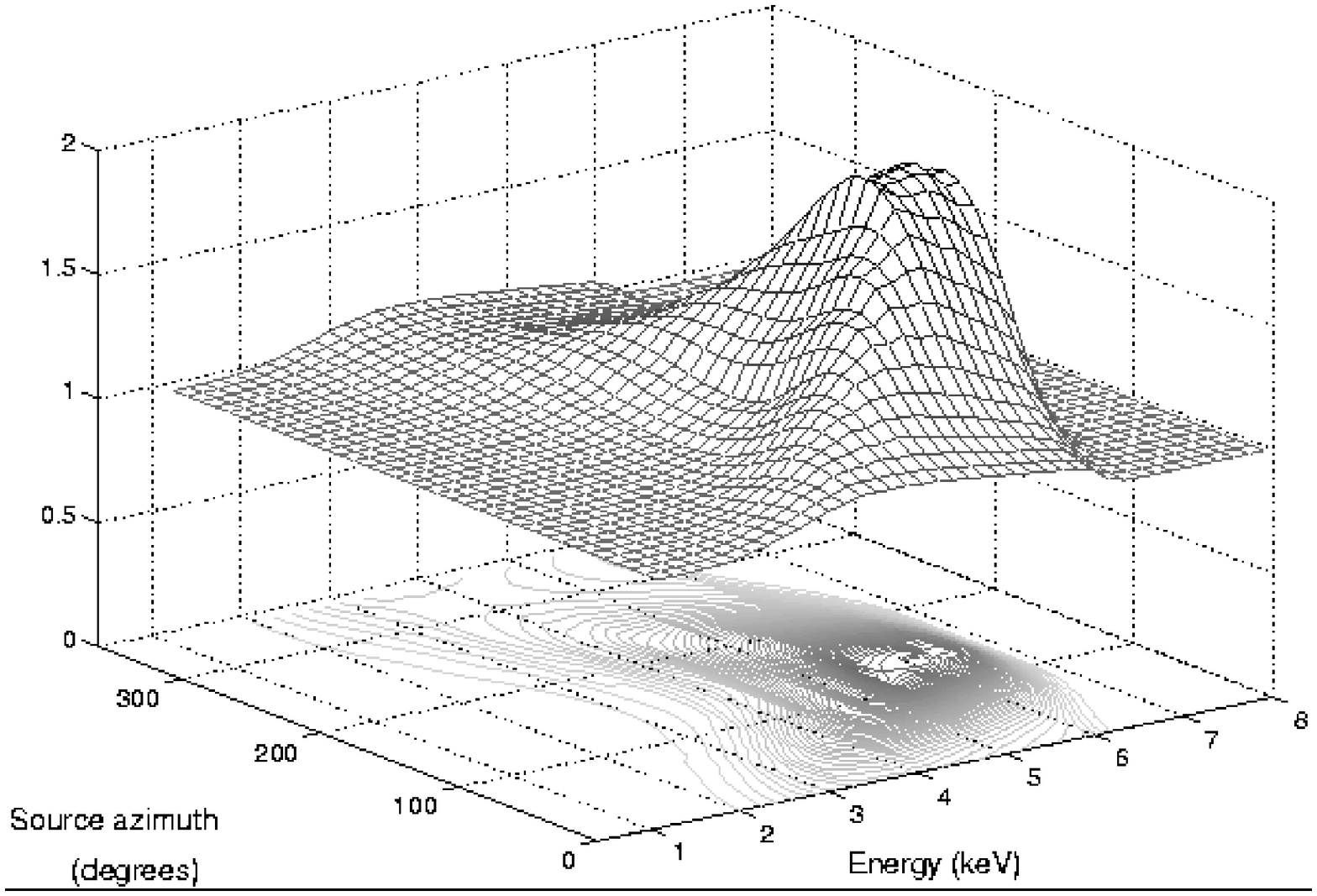}
    \end{minipage}
    }
  \hspace{1cm}
  \subfigure[]{
    \begin{minipage}{8cm}
      \centering
      \label{fig:pointgfrad25}
      \includegraphics[height=8cm,angle=-90]{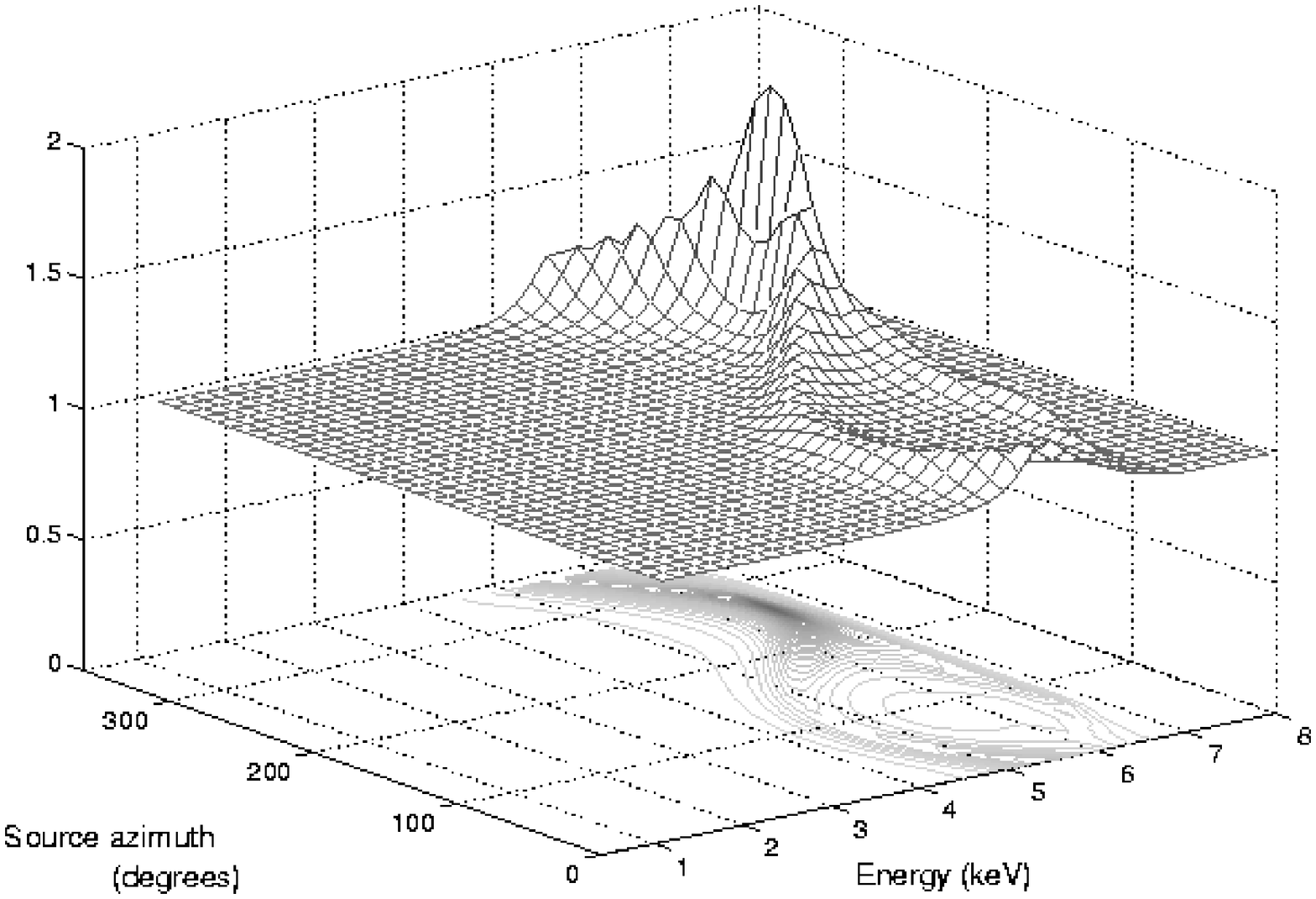}
    \end{minipage}
    }
  \caption{
    \label{fig:transfer-functions}
    (a) Time-dependent spectrum produced by four flares: three
    simultaneous events at $(\rho_s,\phi_s) = (19\,r_g,0^\circ),
    (10\,r_g,100^\circ), (19\,r_g,240^\circ)$ followed by a repeat of
    the latter flare 4$\,t_g (=GM/c^3)$ afterwards. (b) The full `1-d'
    transfer function for time-averaged line profiles, generated from
    azimuthally averaged emissivity distributions corresponding to the
    ring-shaped flares swept out by orbiting flares. (c) Time-averaged
    spectra for a flare at $\rho_s=1.5\,r_g$ for each of the 36
    azimuthal bins in the `2-d' transfer function. (d) Same as (c) but
    for $\rho_s = 10r_g$.  }
\end{figure*}
By following the procedure outlined in the previous section, we
can generate the time-dependent spectrum produced by a single flaring
event at a given location in spacetime. We can now obtain the transfer
function $\clr$ which maps the coordinates of an arbitrary number of
flaring events to the data $\prD$. If the corona itself is optically
thin (so that it has a negligible effect on the iron line emission),
then we can simply superpose the spectrum coming from each individual
flare, accounting for the time delays due to both the different source
positions and the different time at which each flare becomes active.
In this way, the transfer function becomes linear, although this is
not a requirement for the inversion algorithm to work.

In this work we concentrate on the radial and azimuthal locations of
the sources rather than their heights. Rather, we fix the height at a
value which corresponds to a characteristic scale height of the corona
which in practice is set to 2 or 5$\,r_g$, although it would be simple
to extend these models to include the heights of the flares as well.
The time lag arising from a difference in the $t_s$ coordinate of the
source is handled trivially by simply shifting the total spectrum
obtained from a single flare along the time axis by an appropriate
amount. The delays coming from different values of $\rho_s$ and
$\phi_s$ are handled by finding the earliest time at which photons
from a particular source arrive at the observer for each source
position. The entire spectrum is then shifted by this amount in the
full transfer function.  Figure~\ref{fig:timedepgf} shows the
resulting spectrum when the transfer function acts on a set of four
flares, the coordinates of which are given in the caption.
These spectra show the same qualitative features as those in the work of
\cite{1999ApJ...514..164R}, who gave the first examples of such
time-dependent spectra.

While the aim of this paper is to present an inversion method for
reverberation maps, data of this nature is not yet available. Rather,
the best example of a broad iron line to date, seen in MCG--6-30-15,
requires integration times so long that only basic
variability analyses are viable
\citep{1999MNRAS.306L..19I,2001ApJ...548..694V,2002MNRAS.333..687S}.
In this case to get a mean error of $\sim 15\%$, observations of
hundreds of kiloseconds have been necessary
\citep{2002MNRAS.335L...1F}. A recent estimate of the mass of
MCG--6-30-15 gives an upper limit of $\sim 10^7 M_\odot$
\citep{2002MNRAS.329..209M}, corresponding to a characteristic
time-scale ($GM/c^3$) of 49 seconds. For radii in the range
$\sim 2 - 15$, Equation~\ref{eq:orbit_times} implies orbit time-scales of
less than one hour, which means that observations
made with current telescopes such as {\em XMM-Newton} and {\em
  Chandra} span many orbital periods.  This clearly calls for a
modification of the technique as stated above, which we can accomplish
by integrating over both $t_s$ and $\phi_s$ to yield a time-averaged
spectrum for each radius of a flare, modelled as the ring it sweeps
out over time, assuming it persists for an entire orbit. In practice,
we can average the emissivity $F_{out}(r,\phi)$ over $\phi$ to obtain
an axisymmetric profile and use a single time bin extending over the
full duration of flux detection. In this case we can plot the entire
transfer function, as shown in Figure~\ref{fig:ringgf}.

In reality, of course, we do not expect a perfectly axisymmetric
system, due to the turbulent and non-linear nature of the disc and
corona. To begin to probe these effects one can imagine an
intermediate scenario, where a time-averaged spectrum is obtained over
an interval much shorter than the orbit time for a flare, or one where
flaring events occur in `flashes' which persist for times much less
than the orbital period.  In this case, we lose all the temporal
information in the system but we can retain the azimuthal positions of
the sources. The transfer function then consists of a time-averaged
spectrum for each value of the pair $\rho_s,\phi_s$ taken by a flare.
In Figure~\ref{fig:pointgfrad1} we show spectra for a source at a
small radius $\rho_s=2$, for all values of $\phi$.
Figure~\ref{fig:pointgfrad25} shows the equivalent plot for a larger
value of $\rho_s$. Averaging the spectra from these graphs over
$\phi_s$ would produce cross-sections of Figure~\ref{fig:ringgf} at the
corresponding radii.
\section{Computational method}\label{sec:computational-method}
The eventual goal of this work is to construct a framework for
determining patterns of coronal flaring activity from iron K$\alpha$
line spectra.  We can assess the validity of conclusions drawn within
this framework by testing the technique with data generated from a
known distribution of flares.  We operate on this distribution (known
as the `truth') with the transfer function $\clr$ to yield the `mock
data' $\clr\left({\tilde{\prh}}\right)$.  To complete the process we
add a known amount of Gaussian noise $\sigma$, to obtain the simulated
data $\prD = \clr\left({\tilde{\prh}}\right)+\sigma$.
%
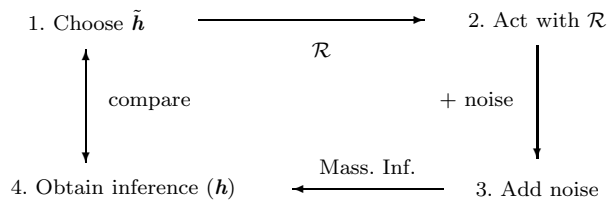
\begin{figure}
  \begin{minipage}{8cm}
    \centering
    \setlength{\unitlength}{1cm}
    \begin{picture}(8,3)
      \put(0,2.25){\makebox(2,1){1. Choose $\tilde{\prh}$}}
      \put(6,2.25){\makebox(2,1){2. Act with $\clr$}}
      \put(6,0){\makebox(2,1){3. Add noise}}
      \put(0,0){\makebox(2,1)[l]{4. Obtain inference ($\prh$)}}
      \put(4,2.25){$\clr$}
      \put(5.7,1.6){$+$ noise}
      \put(4.1,0.7){Mass. Inf.}
      \put(1.3,1.6){compare}
      \put(2.5,2.75){\vector(1,0){3}}
      \put(7,2.4){\vector(0,-1){1.5}}
      \put(5.75,0.5){\vector(-1,0){2}}
      \put(1,1.6){\vector(0,1){0.75}}
      \put(1,1.6){\vector(0,-1){0.75}}
    \end{picture}
  \end{minipage}
\caption{\label{fig:dataprocess} A schematic representation of the process by which we test the
  technique}
\end{figure}
%
\begin{figure*}
\centering
\subfigure[]{
        \begin{minipage}{8cm}
        \centering
        \label{fig:conv_chi2}
        \includegraphics[width=8cm]{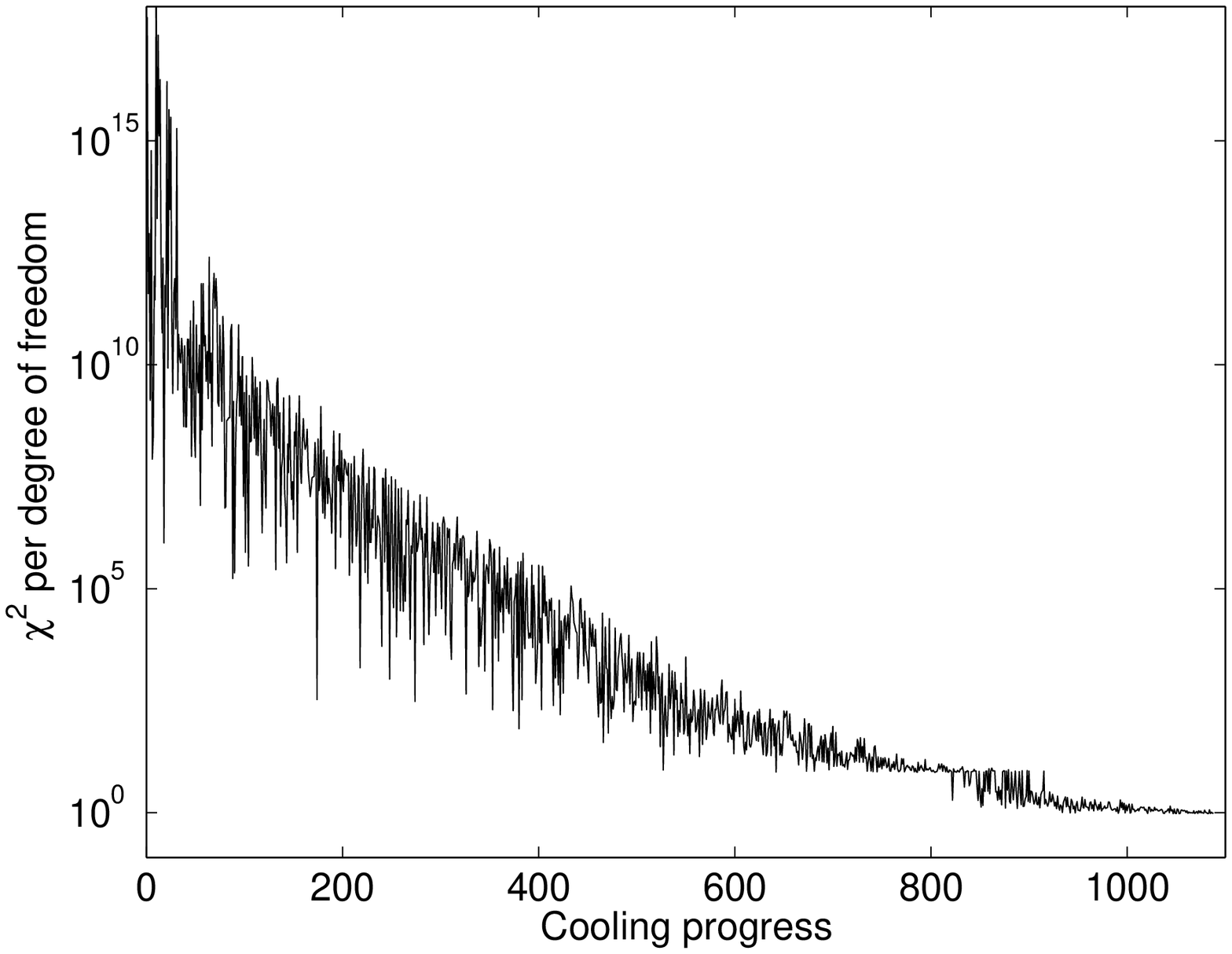}
        \end{minipage}
}
\subfigure[]{
        \begin{minipage}{8cm}
        \centering
        \label{fig:conv}
        \includegraphics[width=6cm]{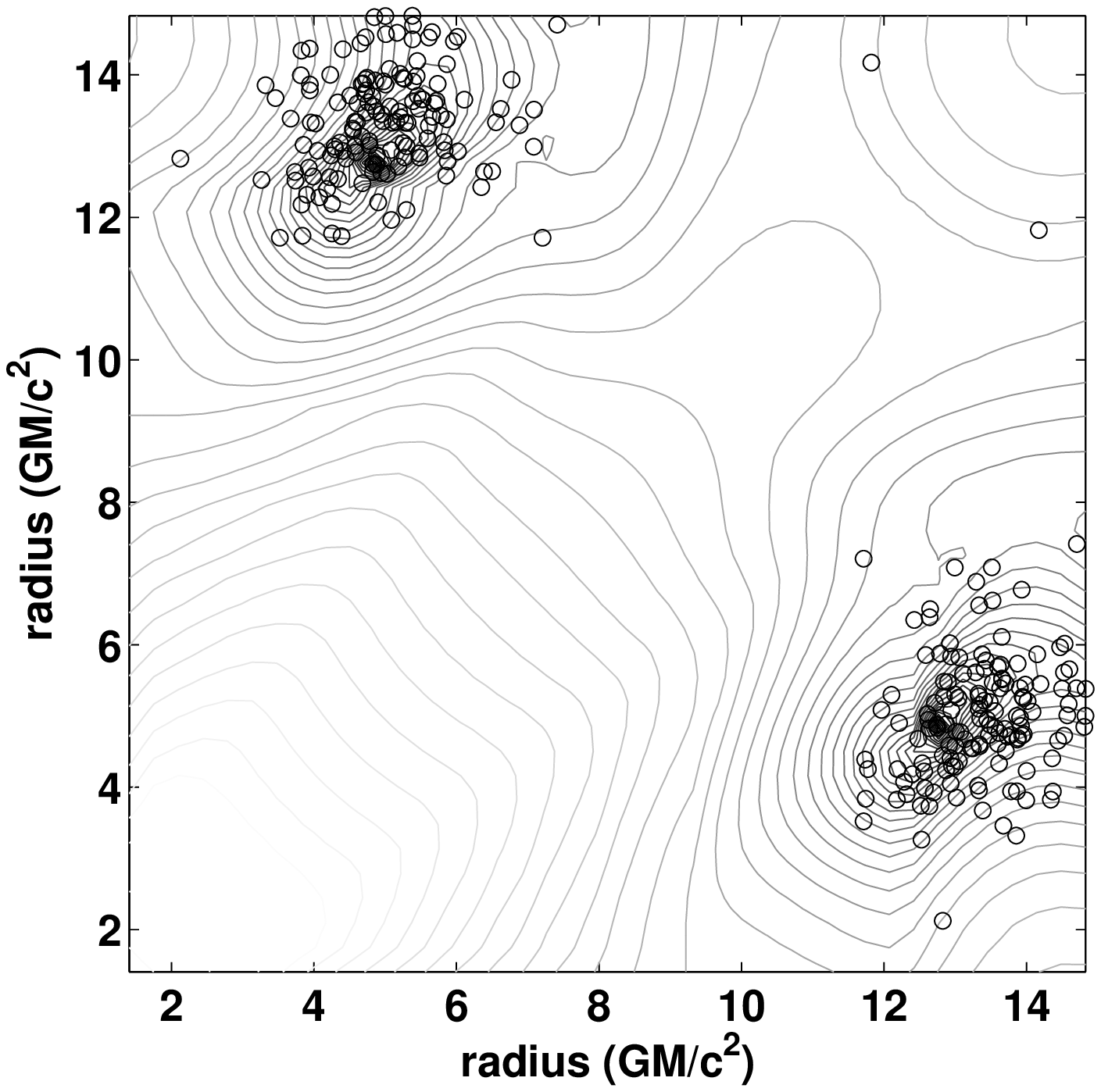}
        \end{minipage} }
\caption{\label{fig:convboth}
  (a) The $\chi^2$ statistic (per degree of freedom) for simulated
  data compared with the mock data produced by operating with the
  transfer function on the mean of the sampled posterior. As the
  algorithm progresses, it can be seen to converge toward the true
  posterior, arriving at a $\chi^2$ of $48.8/50$.  (b) A snapshot of
  some of the final samples. The true posterior is shown as the
  contour plot and the small circles represent the samples, showing that
  they are concentrated in regions of maximum posterior probability.
  The diagonal reflection symmetry is due to the flares being
  indistinguishable. }
\end{figure*}

Given these simulated data we can construct an inference $\prh$ via
the Massive Inference method, which can then be compared with the
truth ${\tilde{\prh}}$. A graphical summary of this process is shown in
Figure~\ref{fig:dataprocess}.

Sampling of the posterior is achieved with a Markov Chain Monte Carlo
algorithm, using a simulated annealing cooling schedule. Each consecutive
sample is chosen according to a `temperature' parameter, which is
increasingly restrictive in terms of the changes it allows as the
system cools. The samples are initially spread over a large region of
parameter space, but as the temperature drops, they become more
concentrated around regions of high posterior probability. The
algorithm can store an ensemble of several flare configurations, each
evolving independently, and the system is cooled as slowly as computation
time allows. In this way, we maximise the number of samples taken and
minimise the chance that the samples are trapped within local minima. 

There is no formal convergence test for the Massive Inference method.
Rather, one can vary both the rate at which the algorithm anneals and
the size of the ensemble of simultaneously evolving parameter sets,
and check that they all converge to a single distribution. A
representation of convergence for a simple example involving only two
flares is shown in Figure~\ref{fig:convboth}. The flares are located
at radii of 5 and 13 $r_g$.  Figure~\ref{fig:conv_chi2} shows the
misfit between the mock and simulated data as the algorithm progresses
from its initial, hot state to the final cool state. The decrease in
temperature is visible as the tapering of the curve's overall envelope
of variation, while the convergence is reflected in the approach of
$\chi^2$ (per degree of freedom) toward the final value of $48.8/50$.
In Figure~\ref{fig:conv} we plot a projection of the sampled posterior
into a plane showing the radial coordinate of each flare, together
with a contour plot of the true posterior, showing that the samples
are indeed concentrated around its peak.
\section{Time-dependent inversions}\label{sec:time-depend-invers}
\begin{figure*}
  \centering
  \subfigure[]{
    \begin{minipage}{8cm}
      \centering
      \label{fig:7flares-xy}
      \includegraphics[height=8cm,angle=-90]{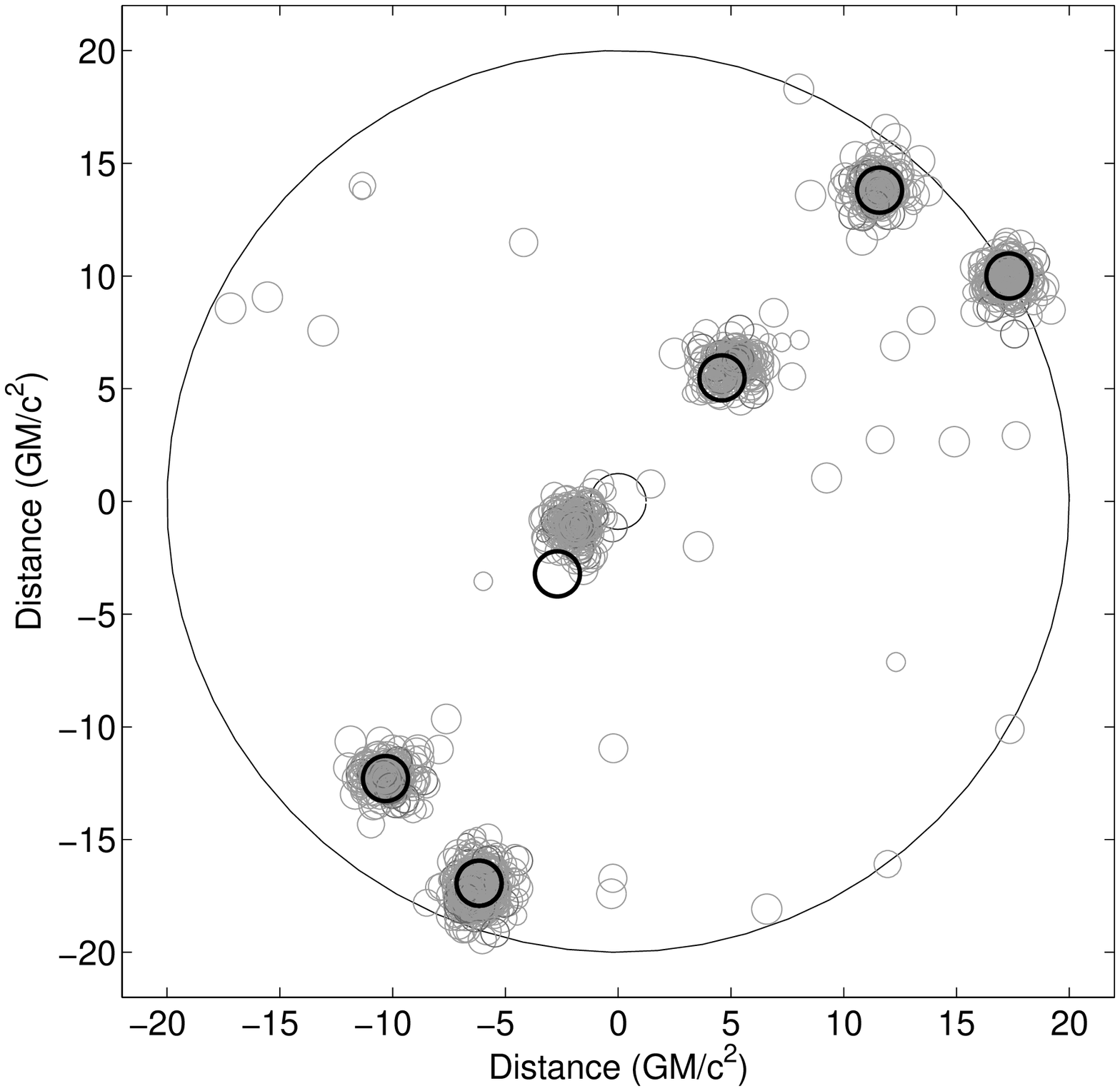}
    \end{minipage}
    }
  \hspace{1cm}
  \subfigure[]{
    \begin{minipage}{8cm}
      \centering
      \label{fig:7flares-ty}
      \includegraphics[height=8cm,angle=-90]{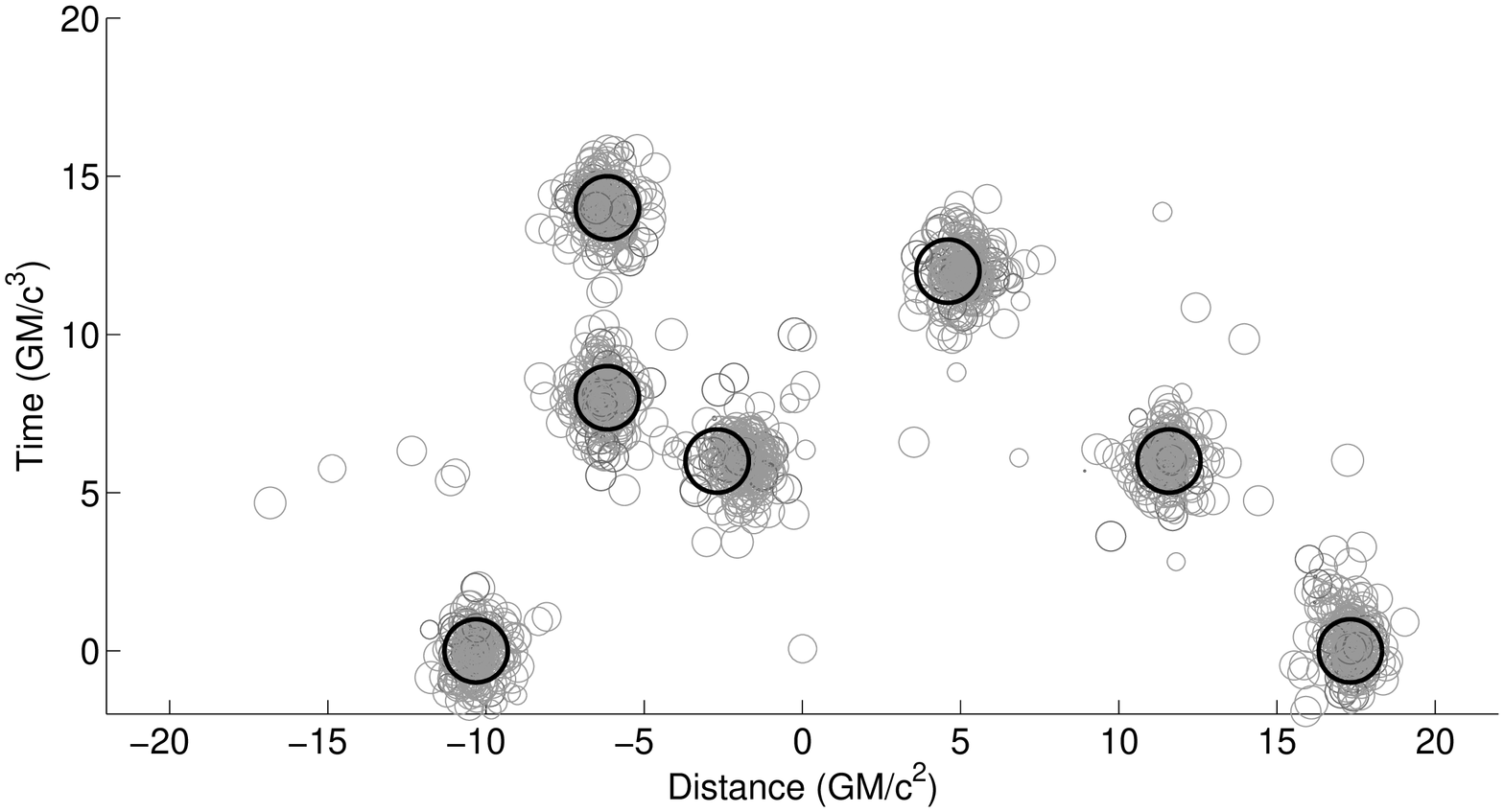}
    \end{minipage}
    }
  \subfigure[]{
    \begin{minipage}{8cm}
      \centering
      \label{fig:7flares-data}
      \includegraphics[height=8cm,angle=-90]{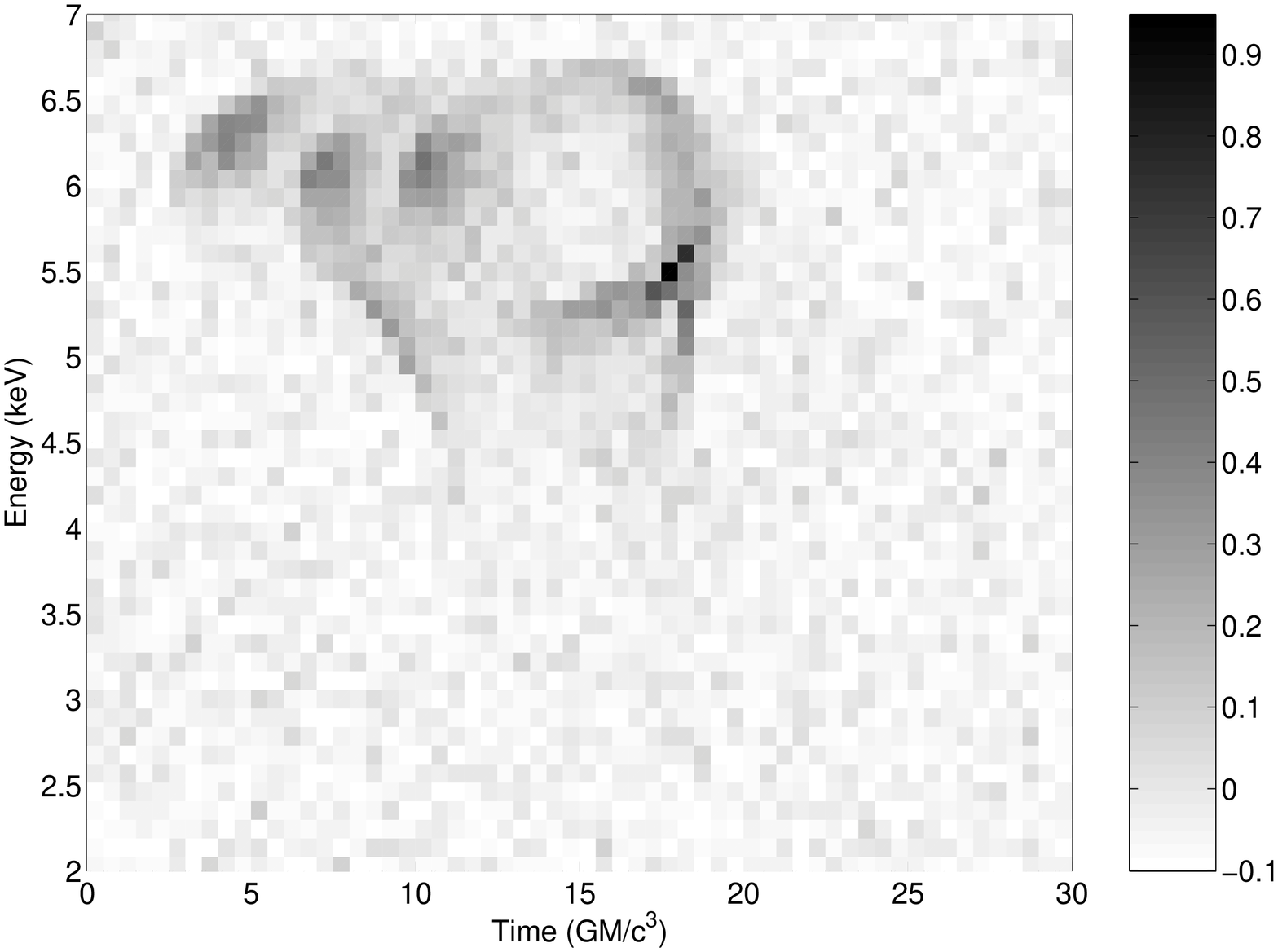}
    \end{minipage}
    }
  \hspace{1cm}
  \subfigure[]{
    \begin{minipage}{8cm}
      \centering
      \label{fig:7flares-hist}
      \includegraphics[height=8cm,angle=-90]{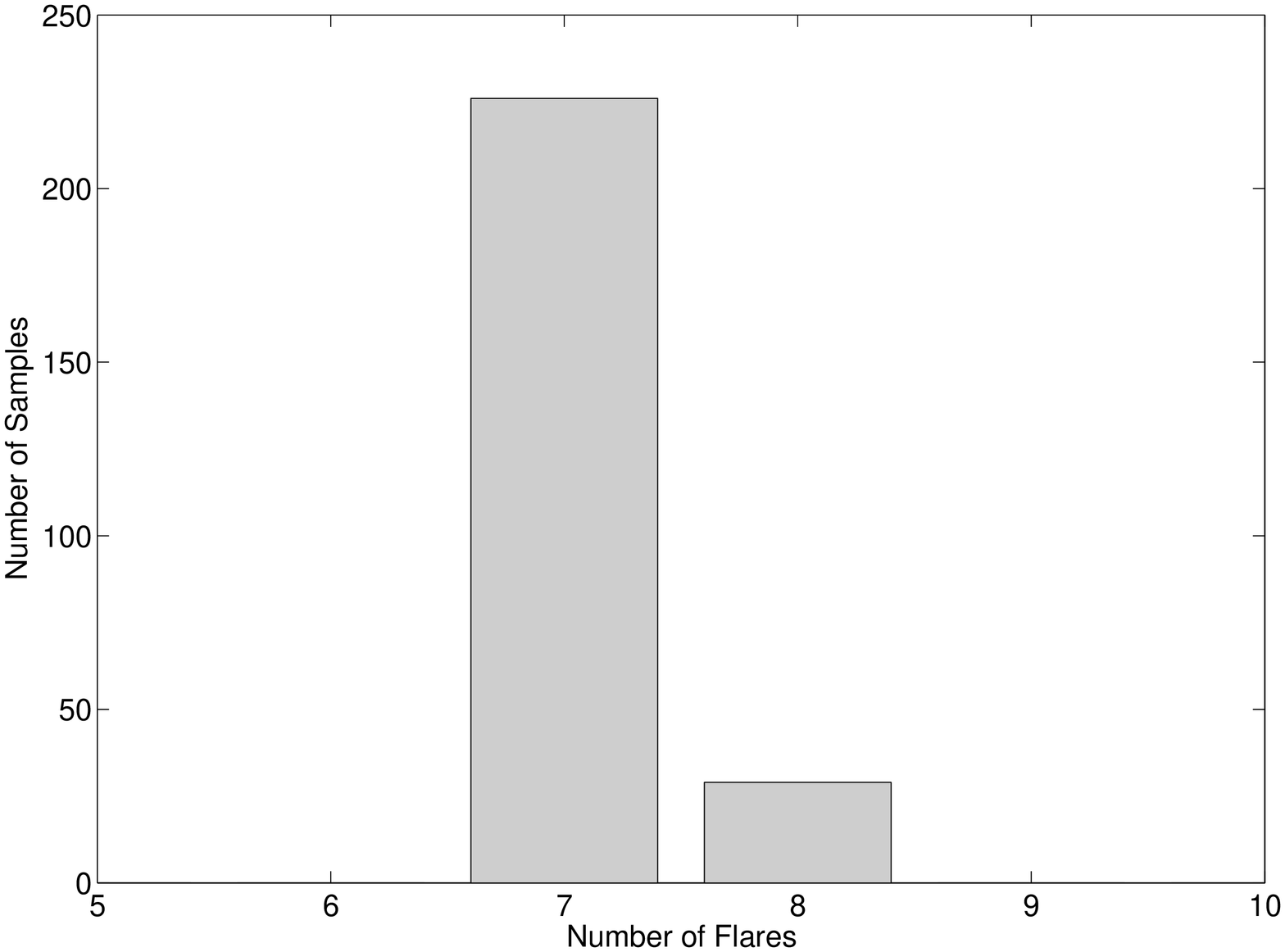}
    \end{minipage}
    }
  \caption{
    \label{fig:7flares}
    (a) Samples from the posterior probability distribution for a
    configuration of 7 flares, projected into the equatorial plane.
    Only 6 flares are visible because two are located above the same
    point in the disc, but are separated in time. The largest circle
    marks the edge of the polar grid of allowed positions, not the edge
    of the accretion disc. (b) The same samples, but now projected
    into the $t$-$x$ plane. The observer is located above the negative
    $y$ axis, nearest the bottom of (a). (c) The simulated noisy data
    from which the inference was made, with a mean SNR = 1. (d) A
    histogram showing the inferred number of flares, $N$.  }
\end{figure*}

We now present some examples of the application of our technique to
simulated reverberation maps, obtained by adding noise to spectra such
as that shown in Figure~\ref{fig:timedepgf}. We quote noise levels as
mean signal-to-noise ratios (SNR), the ratio of the noise to the
average over all pixels in which the flux is non-zero. For a
particular example of simulated data with a mean SNR of 1, which we
take as our starting point for illustrating the technique, the
inferred number of flares is given in Figure~\ref{fig:7flares-hist},
where a strong peak is visible at the true number of 7. The locations
of the flare ensemble above the accretion disc (their $\rho_s,\phi_s$
coordinates) are shown in Figure~\ref{fig:7flares-xy}, where each
sample of the posterior is plotted as a set of $N$ grey circles, where
$N$ is the number of flares corresponding to that particular sample.
In reality, the inner 20$\,r_g$ of the disc are covered by a 20 radii
$\times$ 18 azimuths polar grid of flare positions, but to help
visualise the results, we have displaced each grey circle in a random
direction from the centre of its grid cell by a radius obtained by
sampling from a Gaussian distribution with a standard deviation equal
to the radius of the thicker, black circles, whose centres mark the
location of each flare in the true configuration. The radius of each
grey circle is proportional to the strength of the flare for that
sample and is normalised such that the strongest flare's radius
matches that of the thicker circles marking the truth.

In addition, we plot two circles centred on the origin, which mark the
inner radius of the accretion disc and the radius $\rho_s = 20\,r_g$
which corresponds to the outer limit of the polar grid, but not the
accretion disc itself ($r_{max} = 100\,r_g$). The samples are clustered
around the true positions in all cases but one, showing that the
system copes well with a problem of this complexity, at this level of
noise in the data. Figure~\ref{fig:7flares-ty} shows the inferred
distribution of times at which each flaring event occurred, as a
projection onto the $t$-$x$ plane (the observer is located above the
negative $y$ axis, nearest the bottom of the plot). For the flare
closest to the black hole whose $x$-$y$ position was least accurately
inferred, the samples' time coordinates are centred around the true
value of 6 $t_g$. It is very difficult to pick out 7 distinct spectra
by eye in the simulated data shown in Figure~\ref{fig:7flares-data}, and yet
the algorithm is able to do so, and give the locations of the flares with
a good degree of accuracy.

\begin{figure*}
  \centering
  \subfigure[]{
    \begin{minipage}{12cm}
      \centering
      \label{fig:11flares-3d}
      \includegraphics[height=12cm,angle=-90]{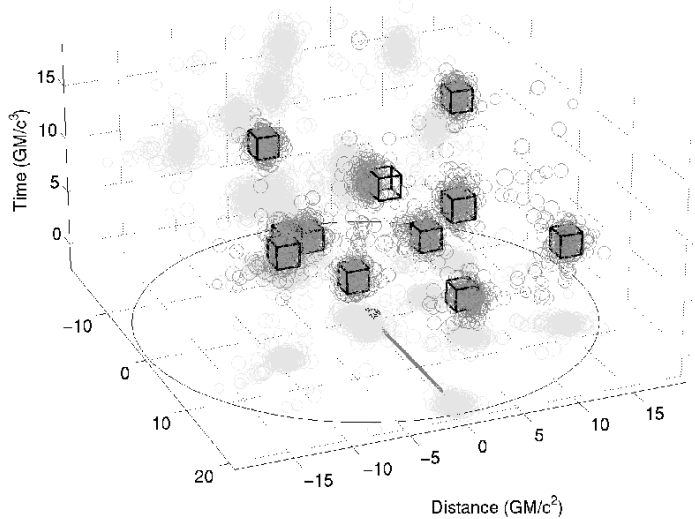}
    \end{minipage}
    }
  \subfigure[]{
    \begin{minipage}{8cm}
      \centering
      \label{fig:11flares-data}
      \includegraphics[height=8cm,angle=-90]{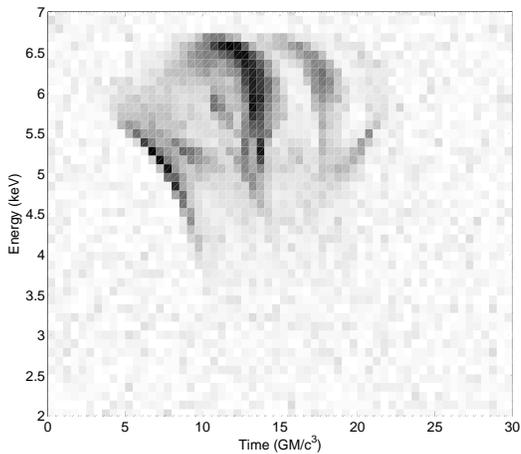}
    \end{minipage}
    }
  \hspace{1cm}
  \subfigure[]{
    \begin{minipage}{8cm}
      \centering
      \label{fig:11flares-hist}
      \includegraphics[height=8cm,angle=-90]{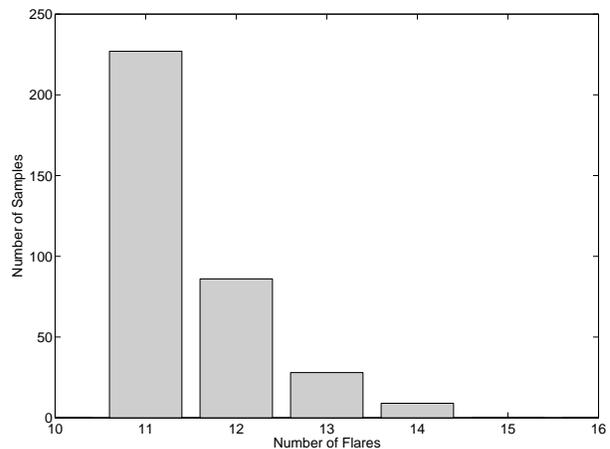}
    \end{minipage}
    }
  \caption{
    \label{fig:11flares}
    (a) Samples from a posterior distribution corresponding to a truth
    consisting of 11 flares, shown as wire-frame cubes. The time at
    which each flare became active is plotted along the vertical axis.
    The height of each flare is fixed at $h_s$ = 5$\,r_g$. The lighter
    circles plotted on the equatorial plane and the vertical plane at
    the back of the plot correspond to the two projections of
    Figures~\ref{fig:7flares-xy} and \ref{fig:7flares-ty} for the case
    of 7 flares. (b) Simulated reverberation map from which this
    inference was made, at a mean SNR = 2. (c) A histogram showing the distribution of
    values of $N$, the number of flares, chosen by the algorithm
    during sampling. }
\end{figure*}

\begin{figure*}
  \centering
  \subfigure[]{
    \begin{minipage}{8cm}
      \centering
      \label{fig:3lares-3d}
      \includegraphics[height=8cm,angle=-90]{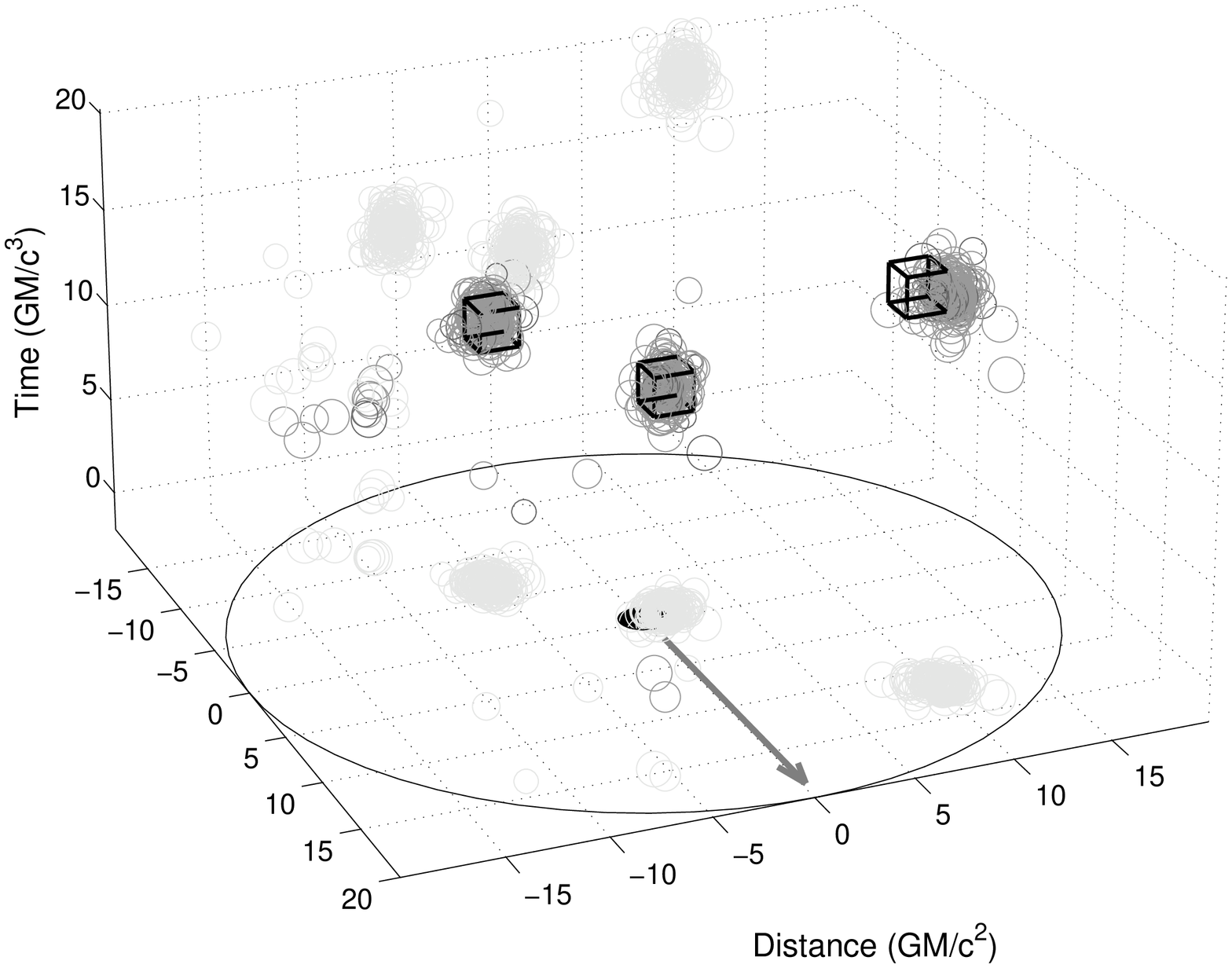}
      \end{minipage}
    }
  \hspace{1cm}
  \subfigure[]{
    \begin{minipage}{8cm}
      \centering
      \label{fig:3flares-data}
      \includegraphics[height=8cm,angle=-90]{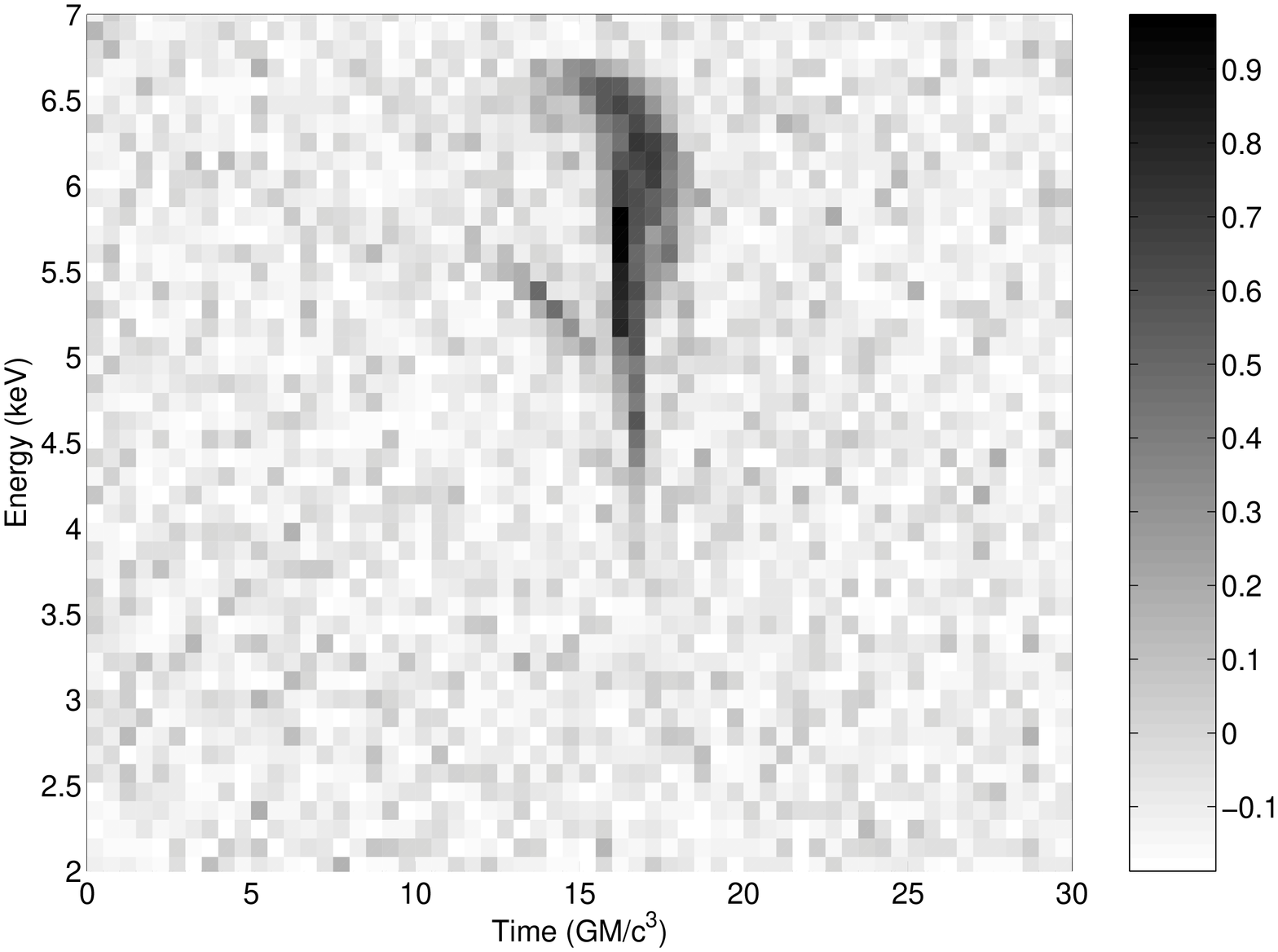}
    \end{minipage}
    }
  \caption{
    \label{fig:3flares}
    (a) Samples from the posterior distribution for a truth with 3
    flares, made at a higher level of noise (mean SNR = 0.5) in the
    simulated data, shown in (b). For this inference, the histogram
    for $N$ (not shown) shows a single dominant peak at the correct
    number of 3 flares.  }
\end{figure*}

We show two more examples in Figures~\ref{fig:11flares} and
\ref{fig:3flares}, one for a configuration of 11 flares and a mean SNR
= 2, and one for poorer quality data (SNR = 0.5) with fewer flares in
the truth.  Figure~\ref{fig:11flares-3d} shows a 3-d representation of
the locations in spacetime of the true and inferred distributions of
flaring events. The vertical axis measures time, not the height of the
flares, which is fixed at 5$\,r_g$. The light grey samples on the disc
and at the side of the plot correspond to the $x$-$y$ and $t$-$x$
projections like those shown in Figures~\ref{fig:7flares-xy} and
\ref{fig:7flares-ty} for the case of seven flares. The truth is
plotted here as a set of wire-frame cubes with a side length equal to
twice the standard deviation of the Gaussian distribution from which
the amount of radial displacement of each dark grey sample was taken.

The noisy data and a histogram of the inferred number of flares are
shown in Figures~\ref{fig:11flares-data} and \ref{fig:11flares-hist}.
It is clear that at least 11 flares were chosen each time in order to
fit the data and higher numbers become increasingly unlikely,
reflecting the prior probability distribution on $N$. Information from
Figures~\ref{fig:11flares-3d} and \ref{fig:11flares-hist} can be
combined by plotting the latter in colour. Instead of grey circles, we
can use a different colour for each value of $N$ in the histogram. We
avoid certain colours being obscured by others plotted over the top by
randomising the order in which they are drawn, thus giving a visual
impression of the histogram, which can be useful in revealing spatial
bias of different values of $N$. Colour versions of all relevant plots
in this paper are available online\footnote{http://www.mrao.cam.ac.uk/$\sim$rg200}.
Figure~\ref{fig:3flares} shows an inversion performed for data at a higher
noise level (mean SNR = 0.5). In both 3-d plots, the thick arrow drawn in the
plane of the disc is directed towards the observer.

As the level of noise is increased, the quality of the inferences
decreases -- flares begin to be missed or show up in incorrect places.
Rather than perform a detailed analysis of the limits of performance
of the above time-dependent setup, we now consider those of
time-averaged spectra, where we focus on just the $\rho_s$ and
$\phi_s$ coordinates of the flares. The reason for this two-fold.
Firstly, the behaviour of the system for demanding noise levels in
time-averaged data is a scenario more likely to be of interest in the
short to medium term, given the quality of current data. Secondly,
obtaining the locations of a set of flares in time is a relatively
easy problem compared to that of revealing their spatial locations,
simply because the structure induced in the data for a single flare at
different times is far more distinctive than at different positions.
It was described in Section~\ref{sec:transfer-functions} that the
former is achieved by simply translating the spectrum from one flare
along the time axis in the data-space by an amount corresponding to
its time-delay, whereas the spectral differences that come from
changes in spatial location are far more subtle. It is therefore
clearer to focus on the time-averaged scenarios described in the
next section.
\section{Time-averaged inversions}\label{sec:time-indep-invers}
We can recast the above system in a form which does not depend on time
by summing the spectra over their entire duration, and removing all
reference to time from the transfer function. In this way, we obtain a
2-dimensional problem, where we aim to recover the $\rho_s$ and
$\phi_s$ coordinates of each source, which take values in the range
1.5 -- 15$\,r_g$ and 0 -- 360$^\circ$ over a grid of 40 radii $\times$ 36
azimuths. In doing this, we have reduced the number of degrees of
freedom (d.o.f.) in the data by almost two orders of magnitude, and as
a result we readily find noise levels at which the system struggles to
produce correct inferences.

\begin{figure*}
  \centering
  \subfigure[]{
    \begin{minipage}{8cm}
      \centering
      \label{fig:point3flares-samples}
      \includegraphics[height=8cm]{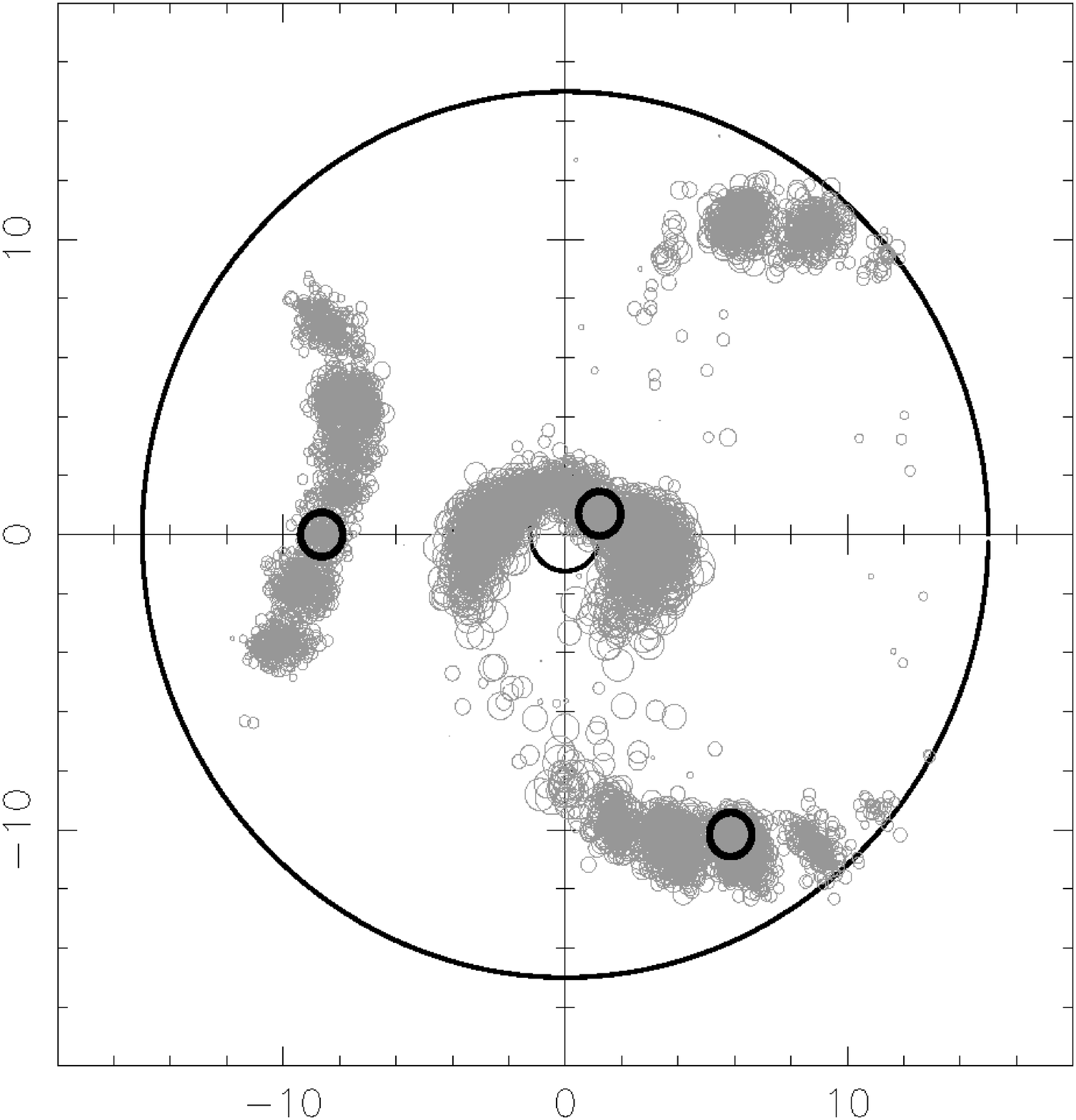}
      \end{minipage}
    }\\
  \subfigure[]{
    \begin{minipage}{7.5cm}
      \centering
      \label{fig:point3flares-data}
      \includegraphics[width=7.5cm]{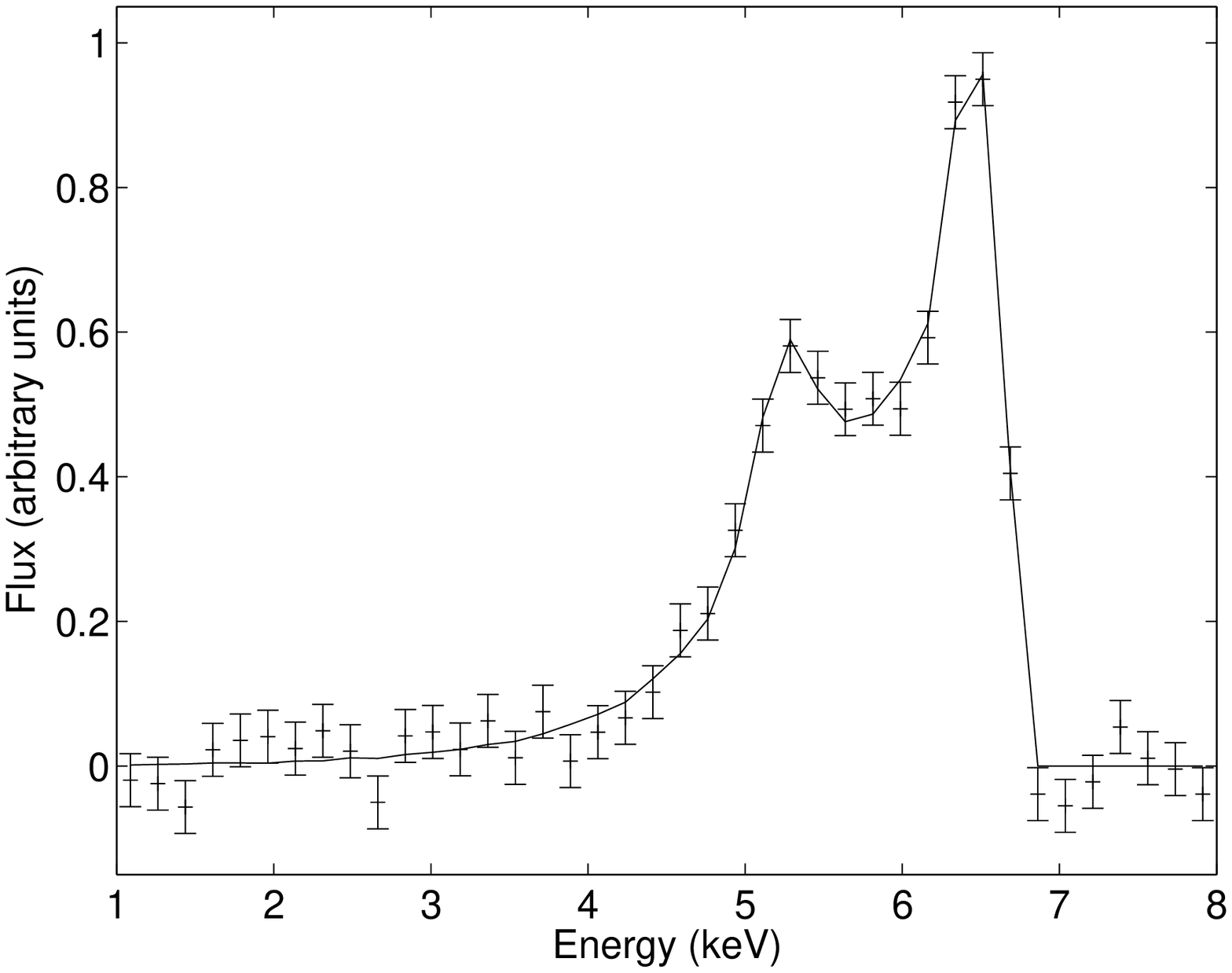}
    \end{minipage}
    }
  \hspace{1cm}
  \subfigure[]{
    \begin{minipage}{8cm}
      \centering
      \label{fig:point3flares-hist}
      \includegraphics[height=8cm,angle=-90]{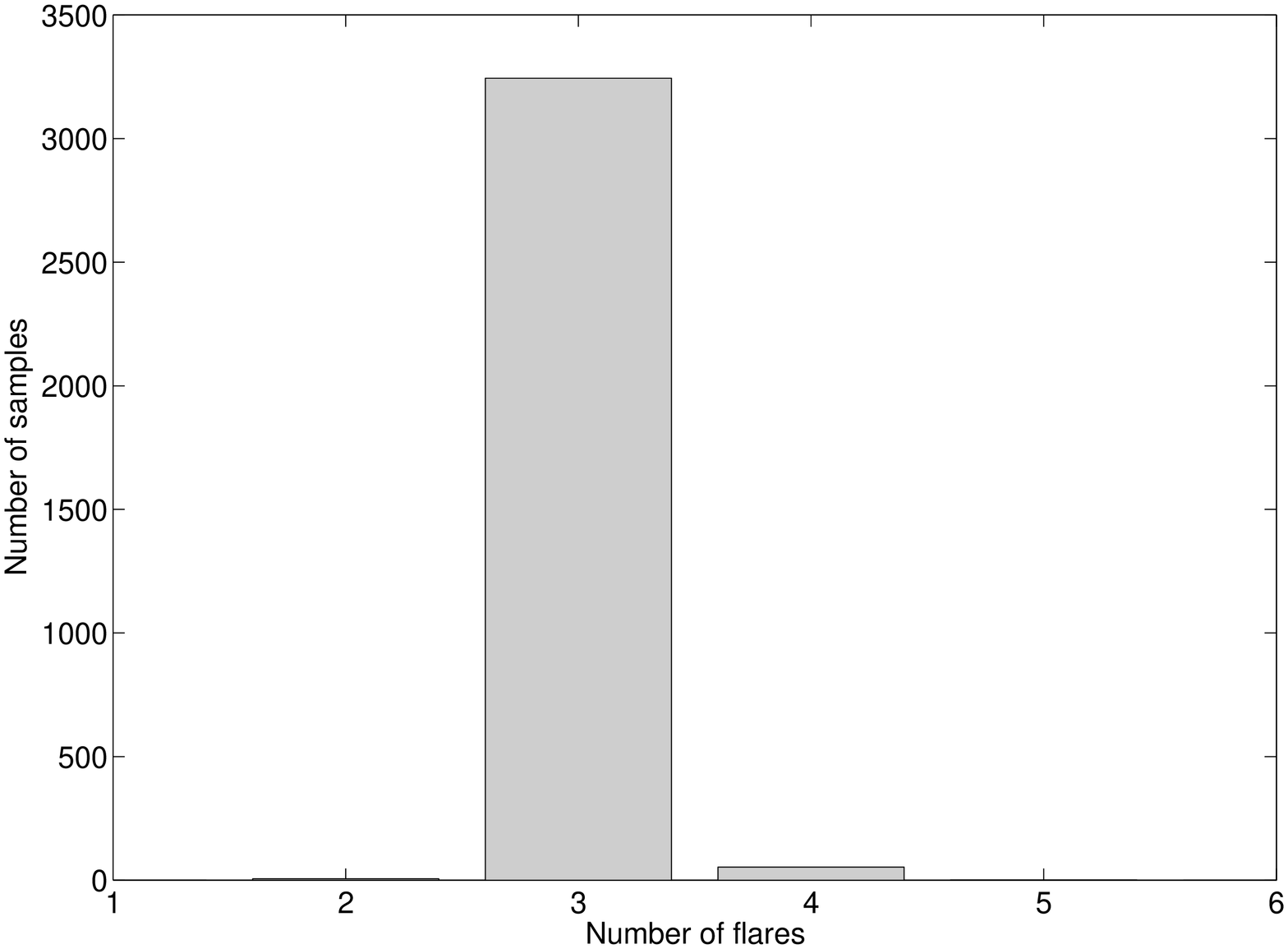}
    \end{minipage}
    }
\caption{
    \label{fig:point3flares}
    (a) Posterior samples for a distribution of 3 flares, inferred
    from time-averaged data, hence only an equatorial plane projection
    is available. The true flare positions are marked with thick
    circles and the large circle at 15$\,r_g$ marks the outer limit of
    allowed flare positions. (b) The time-averaged spectral line used
    in this reconstruction, shown as the points with error-bars. The
    mean signal-to-noise ratio was 7, corresponding to an average
    noise level of $\sim 15\%$. The solid line shows the mock data
    from the inference. (c) A histogram for the value of $N$
    corresponding to each set of samples shown in (a). Although the
    samples are clustered around four distinct regions, the algorithm
    rarely picked sets of four flares, instead preferring the correct
    number of three. }
\end{figure*}

Figure~\ref{fig:point3flares-samples} shows the set of samples of the
posterior probability distribution for an inference based on the data
of Figure~\ref{fig:point3flares-data}. The true distribution contains
three flares at the locations shown by the thick black circles in
Figure~\ref{fig:point3flares-samples}. In addition to being spread out
over many cells of the polar grid of different $\rho_s,\phi_s$ values,
the inference is also clustered around a region above the rear,
receding part of the disc, in which there is no flare in the truth.
The mock data do provide a satisfactory fit to the simulated data,
however, which implies that, at this level of noise, we are probing
degenerate regions of the transfer function.  We note further that the
samples of $N$ shown in Figure~\ref{fig:point3flares-hist} peak very
strongly at the true number of three, so that the extra cluster of
samples in Figure~\ref{fig:point3flares-samples} does not imply the
presence of a fourth flare, rather an additional possibility for the
location of one of the three.
\subsection{Linear-dependence in the transfer function}\label{sec:linear-dependence}
\begin{figure*}
  \centering
  \subfigure[]{
    \begin{minipage}{8cm}
      \centering
      \label{fig:svals}
      \includegraphics[width=8cm]{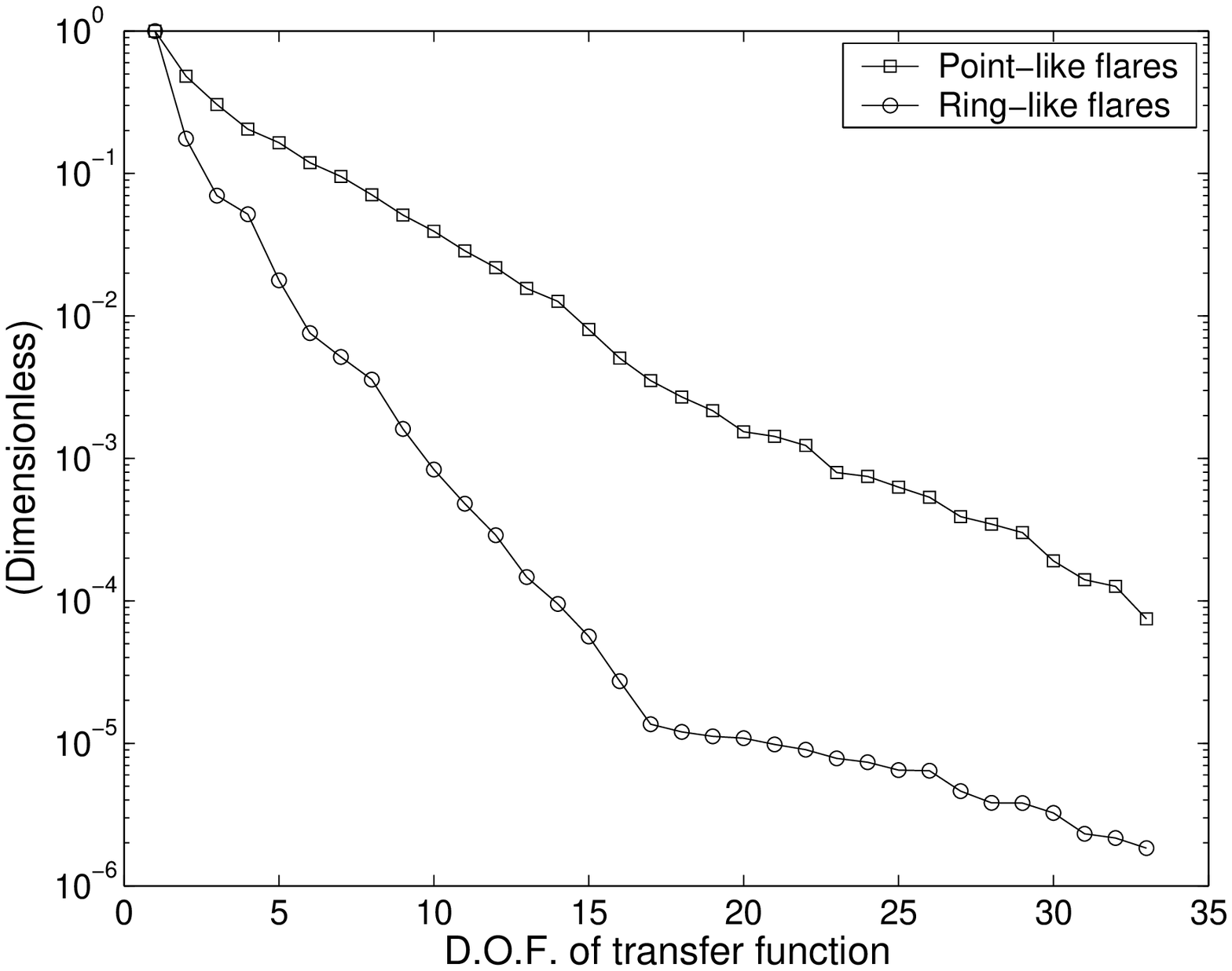}
    \end{minipage}
    }
  \hspace{1cm}
  \subfigure[]{
    \begin{minipage}{7.5cm}
      \centering
      \label{fig:totsvec}
      \includegraphics[height=7.5cm,angle=-90]{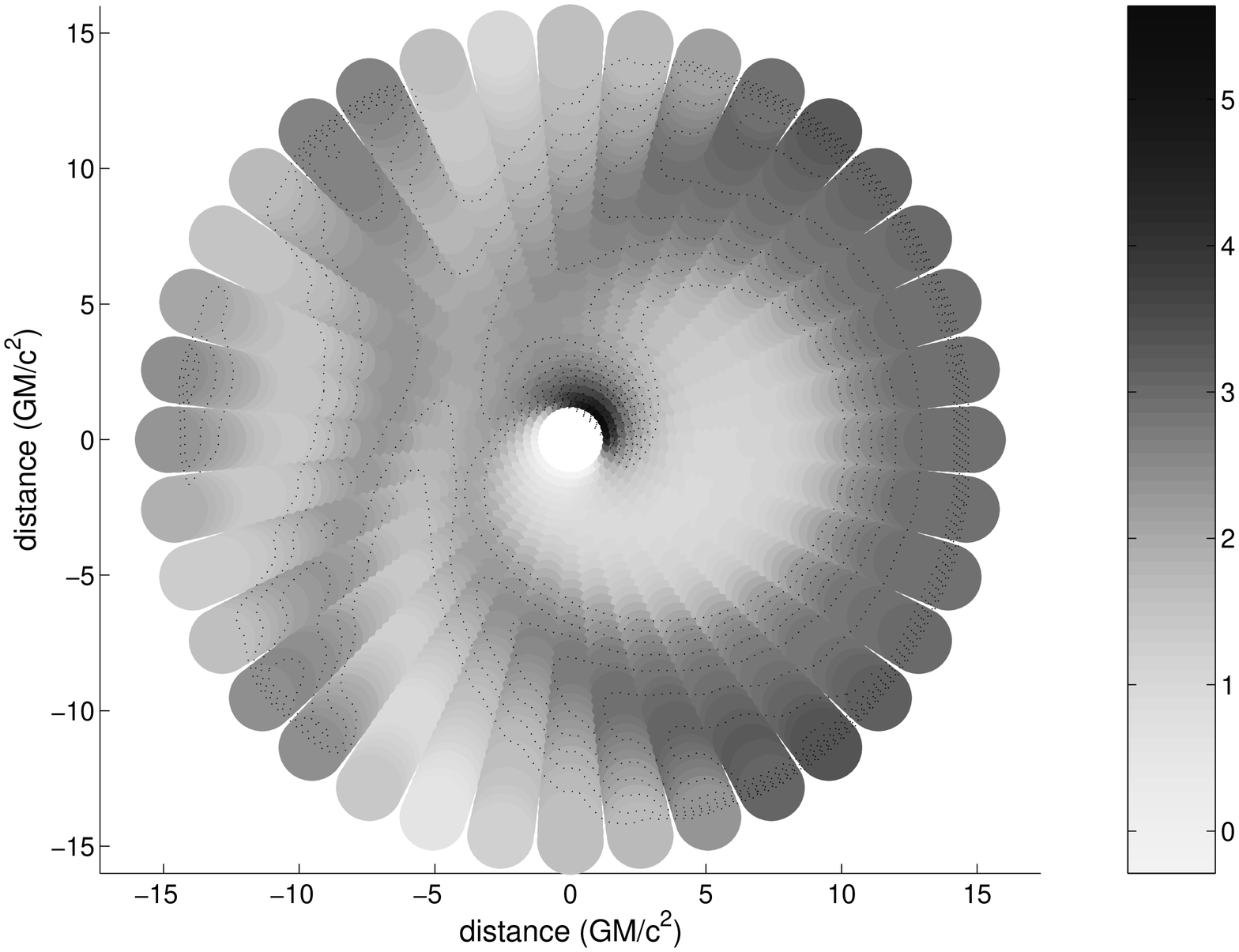}
      \end{minipage}
    }\\
  \subfigure[]{
    \begin{minipage}{7cm}
      \centering
      \label{fig:ring3flares}      
      \includegraphics[height=8cm,angle=-90]{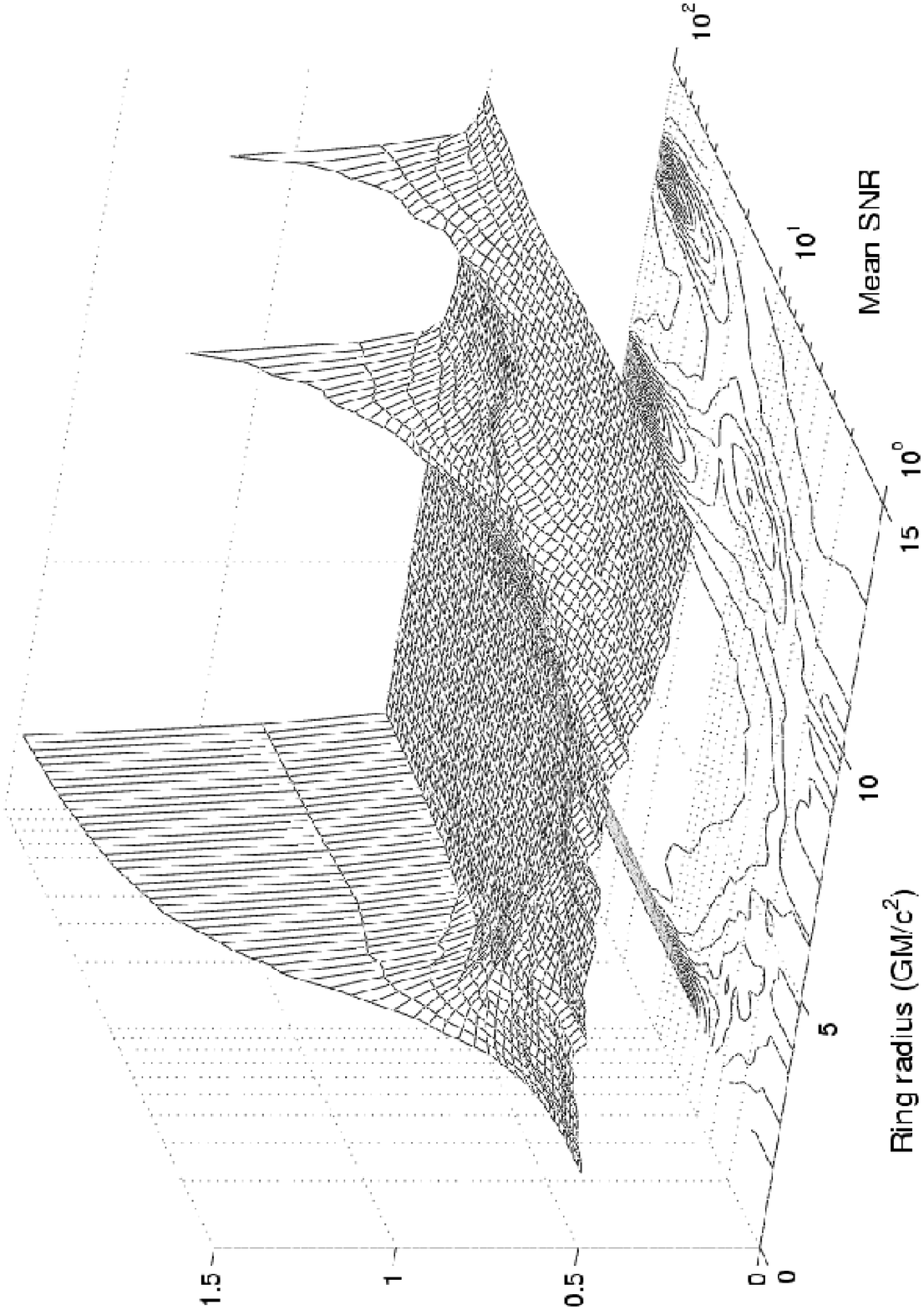}
      \end{minipage}
      }
\caption{
    \label{fig:svd}
    (a) Singular values of the time-averaged transfer functions, for
    the 2-d system (inferring $\rho_s$ and $\phi_s$) shown as squares
    and the 1-d system (inferring just $\rho_s$) shown as circles.
    Losing azimuthal information results in a dramatic reduction in
    the transfer function's linear independence. (b) The average of
    the flare-space singular vectors for the 2-d system, showing
    degenerate regions above the accretion disc, within which a flare
    produces very similar features in the data. On the approach
    (left-hand) side of the disc, beaming produces sharper structure
    in the data which leads to smaller degenerate regions and a
    greater spatial accuracy. (c) Inferred (ring-shaped) flare radii
    and strengths for 3 flares at different levels of noise, plotted
    as the mean flare strength in each of the 40 radial bins. At SNR =
    10, two flaring regions are revealed, but a further factor 10
    reduction in the noise is necessary in order to successfully
    resolve the 3 flares. }
\end{figure*}
We can quantify the linear-dependence present in the transfer function
by performing a singular value decomposition, yielding the singular
values marked by the squares in Figure~\ref{fig:svals}. The gradient
of this line determines the decrease in noise that is necessary, on
average, in order to detect the presence of extra flares as they are
added into the true distribution. The `flare-space' singular vectors
(in the space of all possible flare
configurations) always form a linearly-independent basis, with which
we can represent any flare pattern. However, their relative magnitudes
are given by the singular values, which means that for a given finite
noise level, a certain (large) fraction are effectively zero, removing
this ability to span flare-space and represent an arbitrary pattern.
Degenerate regions arise when the projection of some of a particular
flare pattern lies in these zero directions and fails to result in
detectable features in the data.

A useful measure of the overall effect of this is shown in
Figure~\ref{fig:totsvec}, where we have summed each singular vector,
weighted by the singular values, and projected the result onto the
polar grid covering the inner regions of the accretions disc. A grey
circle is drawn at the centre of each grid cell, whose shading gives
the overlap of the average singular vector with the cell. The radii of
the circles increase linearly with radius to provide a suitable
covering of the disc, for ease of visualisation only. The plot reveals
those cells which, on average, produce the same type of structure in
the data and which require higher levels of noise to distinguish
between them. This explains the patterns seen in the inferred flare
locations shown in Figure~\ref{fig:point3flares-samples} as the
algorithm explores the contours of Figure~\ref{fig:totsvec} with
approximately equal frequency, in its search to fit the data. This
plot also reveals the effective spatial resolution of the technique,
as a function of position on the disc. Two adjacent flares placed in
cells corresponding to a high gradient in Figure~\ref{fig:totsvec}
will be easier to resolve than two placed in a region where the
gradient is small.


As described in Section~\ref{sec:transfer-functions}, we can reduce
the level of complexity in the transfer function still further by
calculating the time-averaged spectra which arise from axisymmetric
distributions of emissivity due to `ring-shaped' flares, modelling
time-averaged point-like flares which persist over an entire orbit or
more. We generate these axisymmetric emissivity profiles by averaging
the illumination pattern due to a single flare over all values of
$\phi_s$, resulting in a 1-d problem where we attempt to recover
just the radius $\rho_s$ of the ring.

Performing this procedure results in a transfer function whose
singular values are given by the circles in Figure~\ref{fig:svals}.
They drop much more quickly than for the 2-d version, giving a
difference of one order of magnitude for just 5 flares. This trend is
reflected in the picture of the 1-d transfer function shown in
Figure~\ref{fig:ringgf}, where the line profile shows only modest
variation as a function of radius. To illustrate the difficulty in
performing inference calculations with this system, we show the
results of many such calculations for a wide range of noise, in
Figure~\ref{fig:ring3flares}, where we plot the mean strength of the
samples in each radial bin. At high noise levels (SNR $\sim$ 1) none
of the three flares present in the truth was located in the inference.
For noise levels comparable to the 2-d calculation of
Figure~\ref{fig:point3flares}, the innermost flare was detected, but
it required a further drop of about a factor of 10 in noise before all
three flares could be detected.
\section{Application to MCG--6-30-15}\label{sec:application-xmm-data}
\begin{figure*}
  \centering
  \subfigure[]{
    \begin{minipage}{6cm}
      \centering
      \label{fig:xmm-samples}
      \includegraphics[height=6cm]{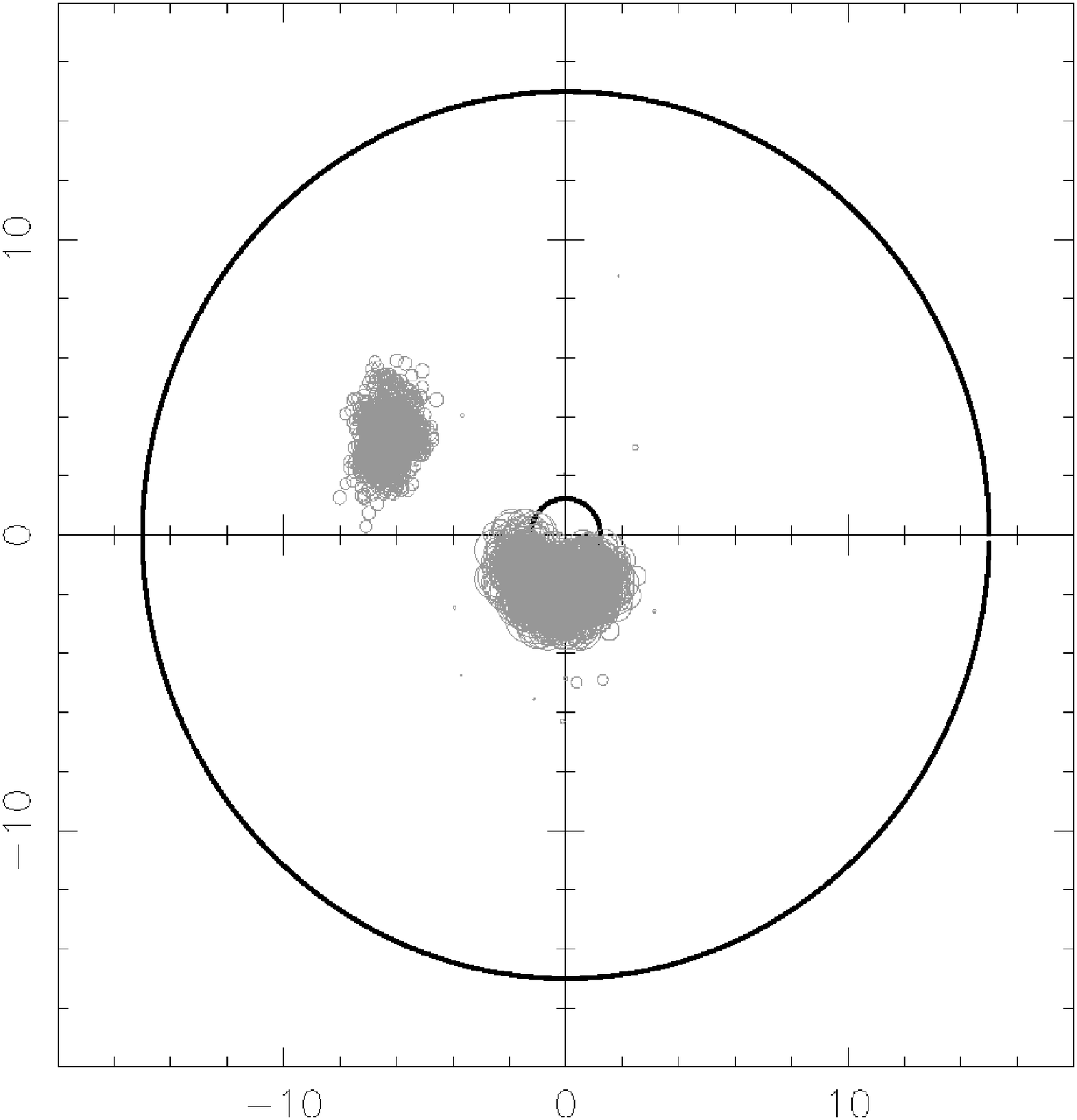}
      \end{minipage}
    }
  \hspace{2cm}
  \subfigure[]{
    \begin{minipage}{8cm}
      \centering
      \label{fig:xmm-data}
      \includegraphics[width=8cm]{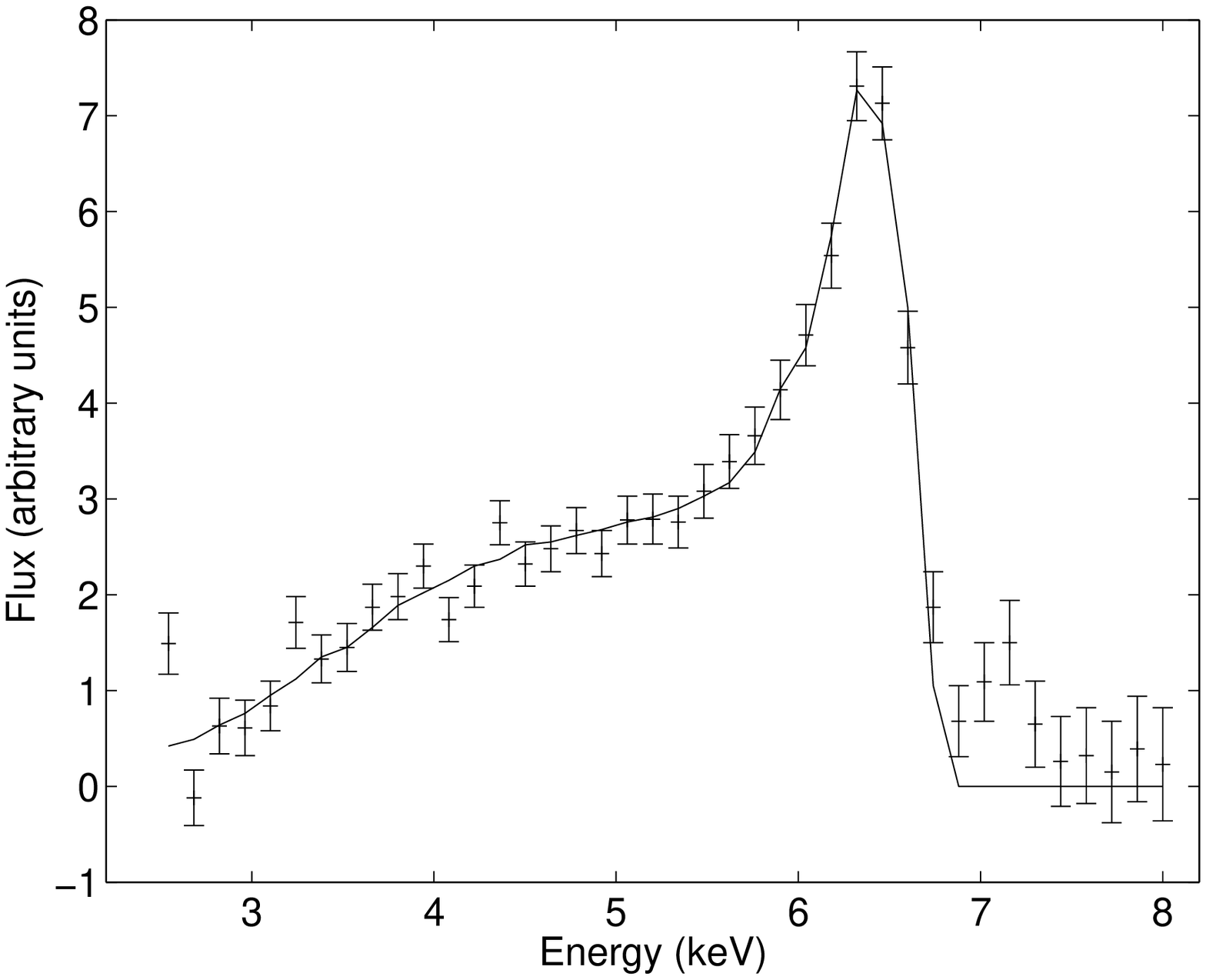}
    \end{minipage}
    }
\caption{
    \label{fig:xmm}
    (a) Inferred distribution of samples for the {\em XMM-Newton} data
    shown as the error-bars in (b). (b) The solid line shows the mock data
    obtained by operating on the inferred flare distribution with the
    transfer function, together with the data themselves.  }
\end{figure*}
We now apply the technique to the results of a 325 ks observation of
MCG--6-30-15 made with {\em XMM-Newton}, described fully in
\cite{2002MNRAS.335L...1F}. We take the spectral line to be the
difference between their Model 4 for the underlying continuum, and the
data. The average noise level is approximately 15\%, which corresponds
to a mean SNR $\sim$ 7. Given the results of
Figure~\ref{fig:ring3flares}, it seems highly doubtful that any
meaningful inferences can be made within the 1-d axisymmetric,
time-averaged framework at the level of noise. We choose instead to apply
the 2-d system,
but with the following large caveat.  Recent estimates of the mass of
MCG--6-30-15 give values close to $10^7 M_\odot$ giving a
characteristic time-scale of 49 seconds, which, as discussed in
Section~\ref{sec:transfer-functions}, corresponds to an orbital period of
less than one hour even for a relatively slow flare, orbiting at
$\rho_s = 15\,r_g$.

It is clear from these time-scales that we cannot hope to attribute
structure in the line profile to particular non-axisymmetric features
in the results of any inference calculation, although such results may
be useful in giving an indication of the general radii at which flares
could be active. In addition, we make no attempt in this work to model
the form of the underlying continuum spectrum, and so are completely
reliant on such a model to define a given spectral line shape. Models
for the continuum, while providing a good overall fit, still allow
considerable variation of the slope in the vicinity of the iron line,
thus altering the shape. Detailed predictions of flaring activity
based on the precise details of this line profile can therefore be
easily changed with a different continuum model, and should not be
taken as clear evidence of a particular flare distribution. This paper
is intended to demonstrate the type of calculation which will become
possible with the advent of high quality, time-dependent spectral
data. The application of our techniques to this {\em XMM-Newton}
observation is a check, to make sure we obtain a plausible picture
from current data which is consistent with previous studies of this
object.

Figure~\ref{fig:xmm-samples} shows the results of applying the 2-d
system to these data. The posterior samples are concentrated in two
regions, and the histogram for the number of flares (not shown) has a
very strong peak at $N=2$. The solid line in Figure~\ref{fig:xmm-data}
shows the mock data from this inferred flare pattern, which provides a
good fit with the data below $\sim 7$ keV. The structure above $\sim
7$ keV is due to iron K$\beta$ emission \citep{2002MNRAS.335L...1F},
which we do not include in our transfer function (although it would be
simple to do so). For this reason, we ignore
the points above 7 keV when calculating the value of $\chi^2$ for this
fit, which is 0.98 per degree of freedom. That the flaring regions are
contained within a few gravitational radii of the central black hole,
which leads to a very steep emissivity profile in this region, is
consistent with the existing picture of this object based on previous
studies.

\begin{figure*}
  \centering
  \subfigure[]{
    \begin{minipage}{6cm}
      \centering
      \label{fig:xmm-truth-good}
      \includegraphics[height=6cm]{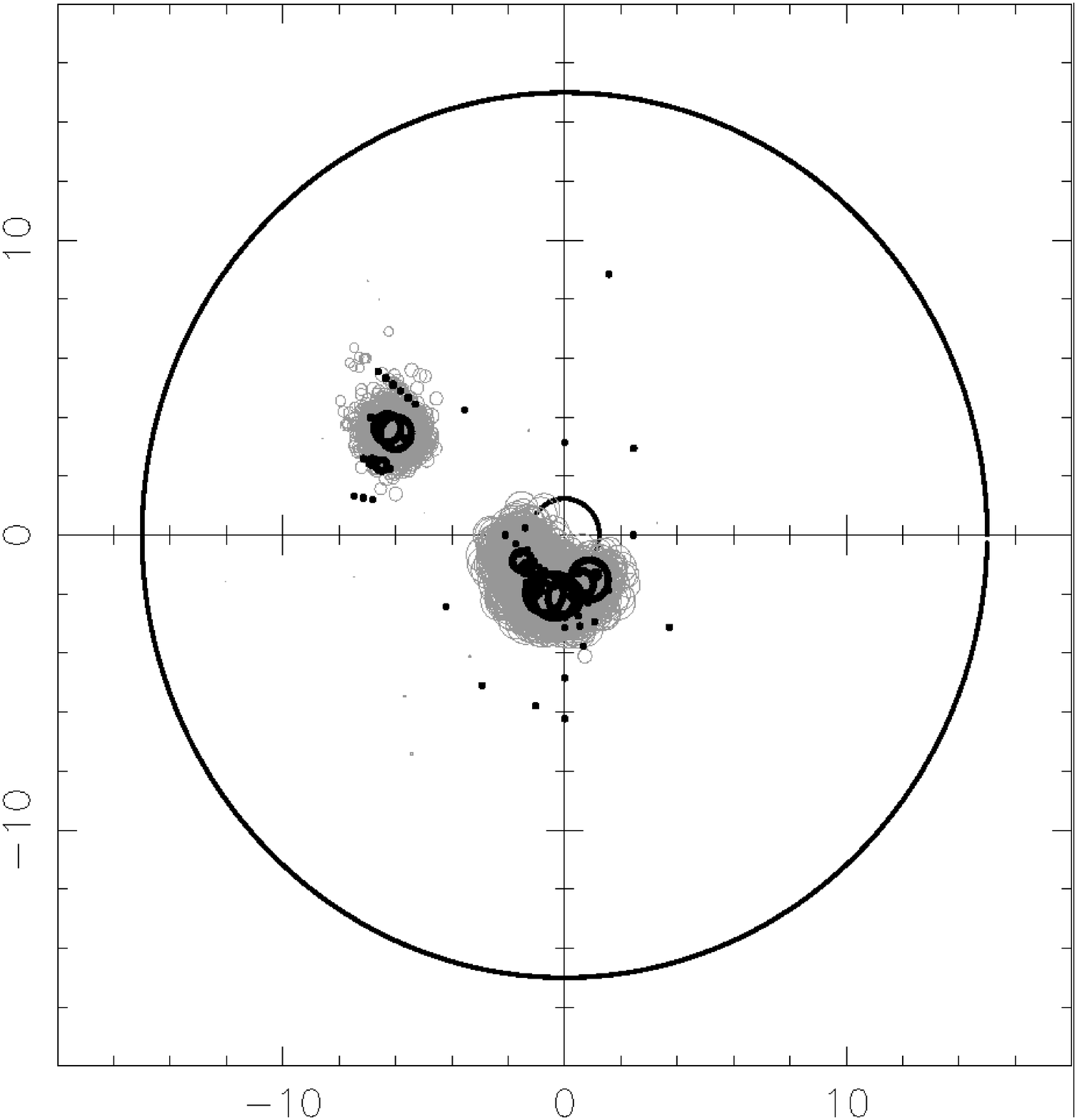}
    \end{minipage}
    }
  \hspace{2cm}
  \subfigure[]{
    \begin{minipage}{6cm}
      \centering
      \label{fig:xmm-truth-bad}
      \includegraphics[height=6cm]{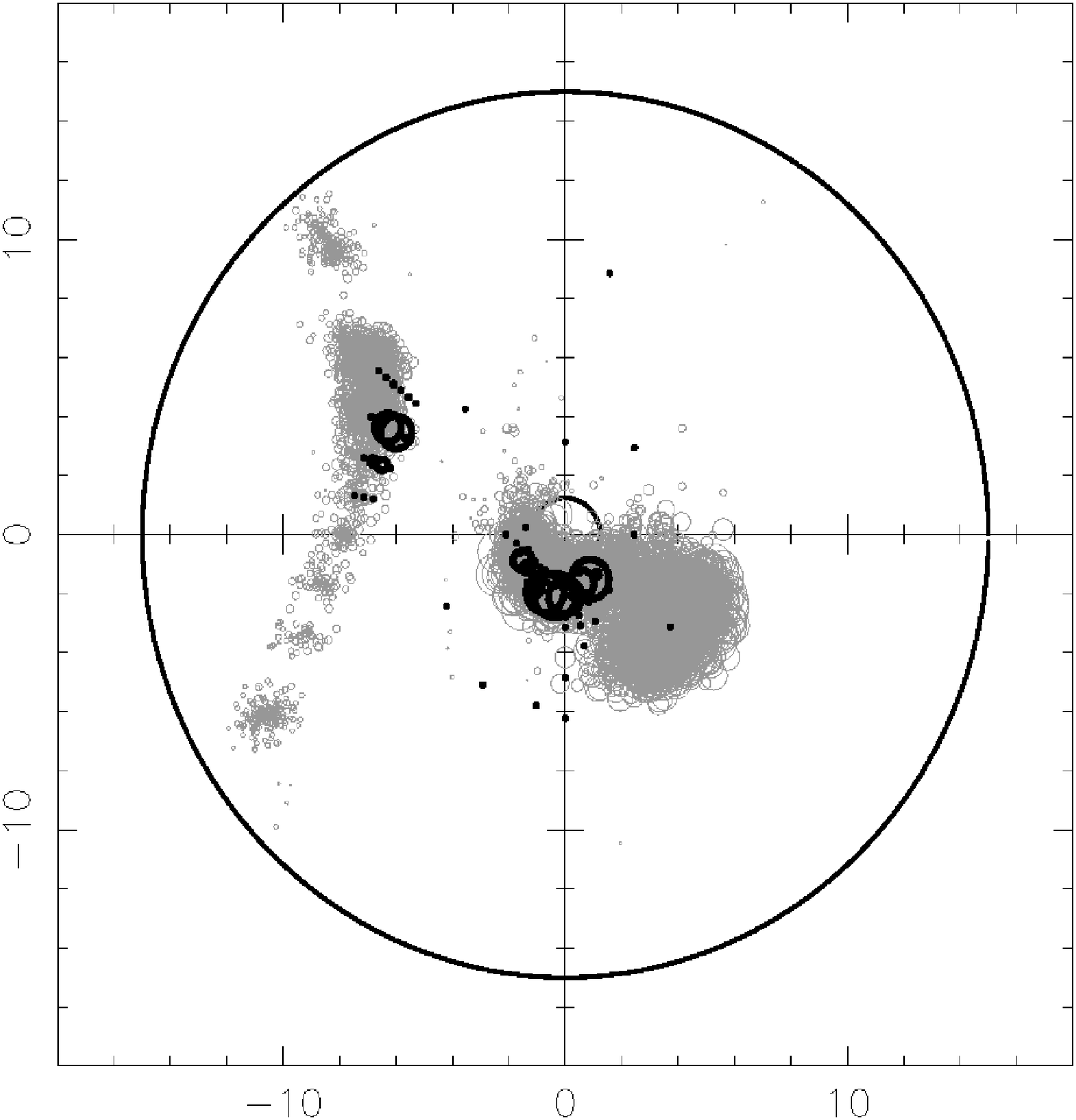}
    \end{minipage}
    }
  \subfigure[]{
    \begin{minipage}{8cm}
      \centering
      \label{fig:xmm-truth-data}
      \includegraphics[width=8cm]{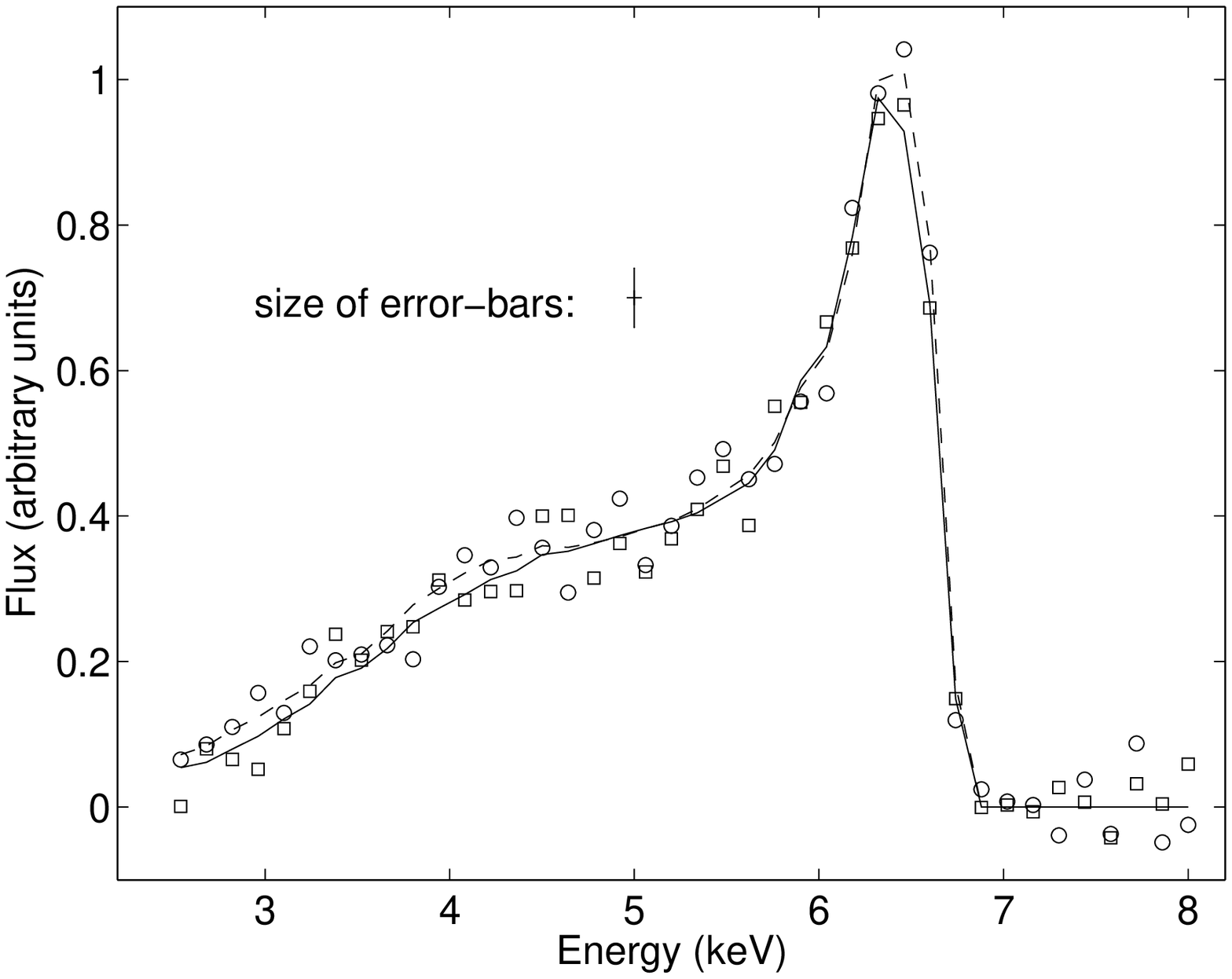}
    \end{minipage}
    }
  \hspace{1cm}
  \subfigure[]{
    \begin{minipage}{7.5cm}
      \centering
      \label{fig:svdproj}
      \includegraphics[width=7.5cm]{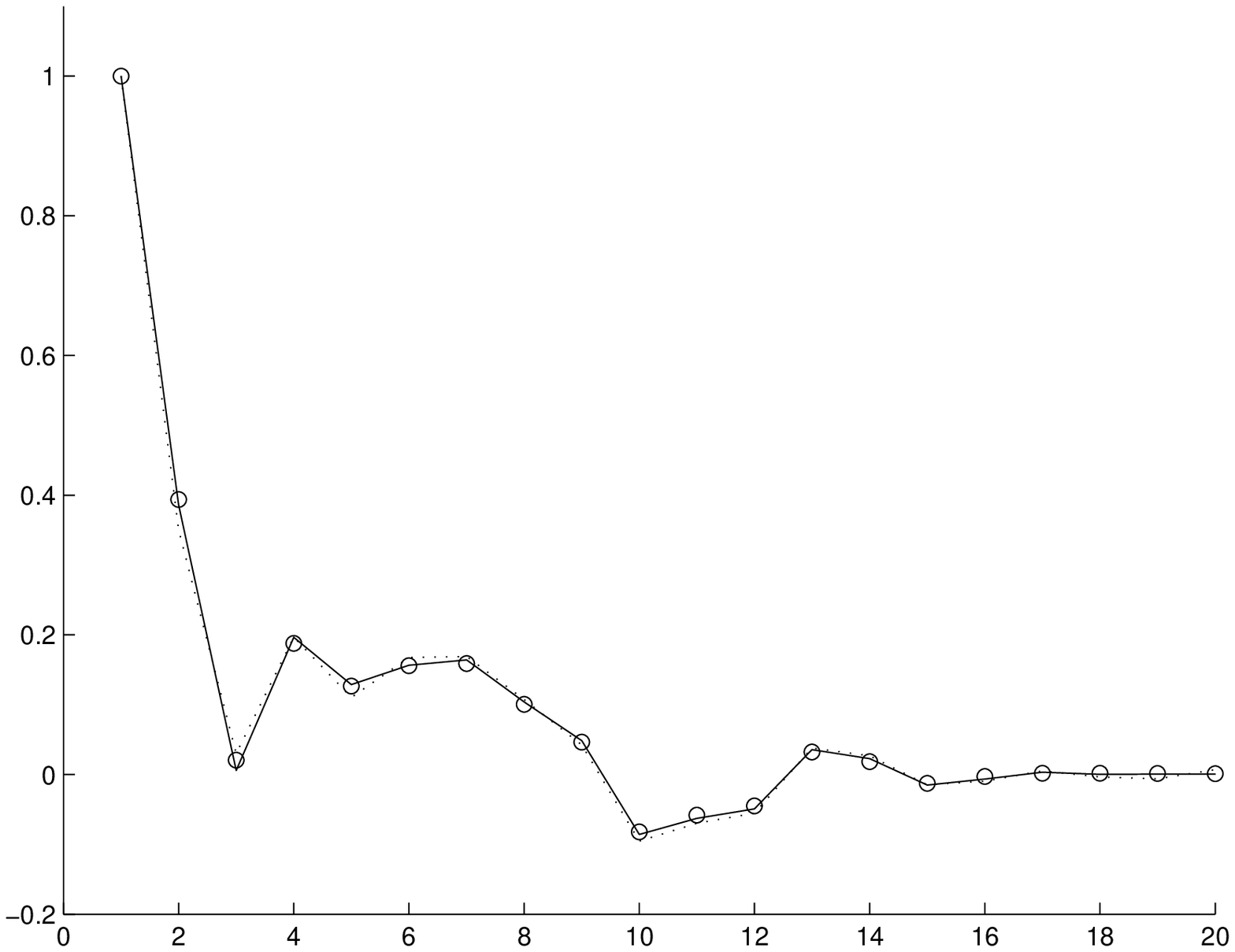}
    \end{minipage}
    }
  \caption{
    \label{fig:xmm-truth}
    (a) Samples from the results of applying the 2-d system to noisy
    data generated from a truth consisting of the results of
    Figure~\ref{fig:xmm-samples}, using a noise level equivalent to
    that in the {\em XMM-Newton} data. (b) Same as (a) except for a
    different random seed used to generate the data. (c) The data
    themselves. The squares show the data used to obtain (a) and the
    circles (b), with the error-bars omitted for clarity.  The solid
    line shows the mock data from the results in (a) and the dashed
    line for (b). (c) The projection of the truth (ie the results of
    Figure~\ref{fig:xmm-samples}) onto the first twenty singular
    vectors, shown as the solid line. The remaining coefficients are
    negligible. The circles show a similar projection for the
    inference in (a) and the dotted line shows the projection of the
    inference in (b).  }
\end{figure*}

Whether the precise azimuthal distribution of flares has a direct
physical significance, given the time-scales involved, is doubtful.
However, it is important to assess the uniqueness of this prediction
-- to what extent would the result change with a different realisation
of the noise in the data? To address this issue, we can perform a
further simulation based on the results of
Figure~\ref{fig:xmm-samples} as follows. By taking a truth consisting
of the mean flare-strength in each grid cell, we can generate mock
data to which we add an amount of noise corresponding to that in the
original data in Figure~\ref{fig:xmm-data}. We can then perform
repeated applications of the inversion technique for different random
seeds in our noise generator. Figure~\ref{fig:xmm-truth} shows the
results of two such calculations. The thick circles show the `truth'
consisting of the results of Figure~\ref{fig:xmm-samples}, and the
grey circles show the samples of the posterior probability. In
Figure~\ref{fig:xmm-truth-good} we obtain very similar results to the
original inference, but in Figure~\ref{fig:xmm-truth-bad} we see some
differences. Again, we can interpret this result within the singular
vectors of Figure~\ref{fig:totsvec} -- the `erroneous' samples are
continuously connected along regions of high degeneracy.  For this
reason, we should interpret the results of
Figure~\ref{fig:xmm-samples} as indicating the presence of flares
anywhere within the two regions in Figure~\ref{fig:xmm-truth-bad}.
Figure~\ref{fig:xmm-truth-data} shows the noisy data for the two
different realisations of noise and the solid and dashed lines are the
mock data corresponding to Figures~\ref{fig:xmm-truth-good} and
\ref{fig:xmm-truth-bad} respectively. Another way of seeing how similar
the two different flaring patterns are, when considered in the context
of the singular vectors, is to evaluate the components of the two
patterns on the singular vector basis. The solid line in
Figure~\ref{fig:svdproj} gives the projection of the `real' results of
Figure~\ref{fig:xmm-samples} onto the singular vectors, and the
circles and dotted line correspond to those for
Figures~\ref{fig:xmm-truth-good} and \ref{fig:xmm-truth-bad}
respectively. It is through projections like these that we can assess
the true similarity of structure in the data that different patterns
of flaring activity can produce.
\section{Conclusion}\label{sec:conclusion}
We have presented a new way of analysing line profile data which
constrains the number, positions and strengths of hard X-ray sources
in an accretion disc's corona. The method makes use of the `Massive
Inference' prior probability distribution which is based on an
assumption of discrete, point-like sources.

By applying it to simulated noisy reverberation maps, we found that
the technique performs well, making the most of the rich information
content of such data. The algorithm was able to disentangle the
complicated signatures of combinations of flare events, rather than
concentrating on scenarios where it can be assumed that one or two
dominant flares are responsible for the observed data
\citep{1999ApJ...514..164R,2000ApJ...529..101Y}. Indeed, if there is
only a single large flare present, such a conclusion will be reached
by our method.

Reverberation mapping of relativistic spectral lines is still some way
off, and so we modified our system to treat time-averaged spectra at
similar levels of noise to those obtained by current instruments such
as {\em XMM-Newton} and {\em Chandra}. In doing so we lost all the
temporal information in the transfer function and reduced the problem
to that of inferring the set of radial and azimuthal coordinates of
the flares. The relative information loss in the data, however, was
greater than in the parameter space, and the resulting inference
problem was more challenging. At realistic levels of noise, we found
some ambiguity in the inferred spatial locations of the flare, but
less in the inferred number present in the ensemble. This ambiguity
was quantified using a singular value decomposition, which yielded a
`map' of the disc showing the relative ease with which different
regions of the disc could be differentiated, and gave a limit on the
spatial resolution of the technique.

We reduced the problem still further, considering data collected on
time-scales greater than the orbital period of each flare. In this
case we averaged the illumination pattern of a flare over the
azimuthal coordinate $\phi_s$ to obtain a 1-d transfer function based
on `ring-like' sources, and attempted to constrain the radii of these
rings. This problem proved to be very difficult, requiring very low
levels of noise to obtain useful results. This very high level of
degeneracy in the transfer function is due to the line profile being
affected only weakly by the position of the source, the dominant
contribution coming from the relativistic effects occurring during the
photons' subsequent journey from disc to observer.

As a consistency check, we applied the technique to a recent observation of the iron
K$\alpha$ line in MCG--6-30-15 made by {\em XMM-Newton} over 325 ks.
This long integration time means that, given the mass of the central
black hole of $\sim 10^7 M_\odot$, it is not possible to resolve any
structure in the data due to non-axisymmetric illumination of the
disc, and that the 1-d ring-based transfer function is most
appropriate. However, given the difficulties present in the highly degenerate ring-based
system described above, we applied the 2-d time-averaged algorithm to
this data, allowing some azimuthal structure in the corona. We found
two regions of flaring activity, one at a radius of $\sim 2\,r_g$ on
the near side of the disc which is responsible for the bulk of the
highly redshifted emission below 5 keV, and one on the approaching
side at $\sim 6\,r_g$ which is more beamed and produces the tall peak
close to 6.4 keV. 

Further simulations for different realisations of
random noise illustrated explicitly that this result must be
interpreted with the singular vector map of Figure~\ref{fig:totsvec}
in mind, spreading out the regions occupied by the samples to cover those
in Figure~\ref{fig:xmm-truth-bad}. This approach showed that, within the
assumptions underlying the method (mainly that different locations of
X-ray flares in the strong-gravity region of the central black hole account for
structure in the line profile), the results of Figure~\ref{fig:xmm-samples}
are quite robust. Physically, however, we should not expect to be able to
pick out any particular azimuthal region based on a 325 ks observation.
In fact, by including a relatively weak, narrow component in the
line in order to account for the tall peak near 6.4 keV, it is possible to
achieve a very good fit with the data based on an azimuthally averaged illumination pattern
\citep{miniutti_et_al_2003}. In addition, the uncertainty that comes from defining a
spectral line shape relative to a given model for the underlying
continuum radiation, means that the above results (other than the
loose condition that the flares are contained within $\sim 15\,r_g$)
can be changed very easily by making different assumptions about the
continuum. For these reasons, we do not claim that the results of Figure~\ref{fig:xmm-samples}
are a meaningful prediction for MCG--6-30-15, rather, it is reassuring to
obtain a pattern contained within a few gravitational radii of the central
black hole, and gives an indication of the sort of results which may become
viable with higher quality data. It is with the advent of high-quality, time-dependent spectral
data, like that promised by missions such as {\em Constellation-X},
that inversion techniques such as those described here offer the
possibility of mapping the hard X-ray emission within a few
gravitational radii of a black hole.
\section*{Acknowledgments}
The authors are grateful to Andy Fabian for providing the data used in
this research. RG is grateful to Steve Gull, John Skilling and
Giovanni Miniutti for some very useful discussions, and was supported
by a PPARC studentship.
\appendix
\onecolumn
\section{Geodesic motion}\label{sec:form-geod-moti}
The gravitational calculations in this paper were performed using
geometric algebra, within a new formulation of gravity theory recently
developed by \cite{DGL98-grav}.  In this appendix we translate some of
the results of these calculations into conventional notation, but
mention only in passing the motivation for casting the problem in this
form.  For a detailed account of the tools and ideas which lead to
these results, see \cite{DGL98-grav}, \cite{hes-sta} and \cite{hes-gc}.

We begin with an orthonormal tetrad with components on the
Boyer-Lindquist coordinate frame given by
\begin{equation}\label{sphbasis}
  \that  =  \frac{r^2+a^2}{\rho\sDelta}\,\ddt +
  \frac{a}{\rho\sDelta}\,\ddphi \qquad \quad
  \rhat  =  \frac{\sDelta}{\rho}\,\ddr \qquad\quad
  \thetahat  =  \frac{1}{\rho}\,\ddtheta \qquad\quad
  \phihat  =  \frac{a\sin\theta}{\rho}\,\ddt + \frac{1}{\rho\sin\theta}\,\ddphi
\end{equation}
where
\begin{equation}\label{rho_and_Delta}
  \rho^2 = r^2+a^2\cos^2\theta,\qquad \Delta = r^2 + a^2 - 2r.
\end{equation}
We can then parametrise a photon's momentum with the direction-cosines
$\alpha$ and $\beta$ and the energy parameter $\Phi$ as follows
\begin{equation}\label{bl_phot_mom_expanded}
  p = \Phi\,\left(\that + \cos\alpha\cos\beta\,\rhat -
    \sin\alpha\cos\beta\,\thetahat + \sin\beta\,\phihat \right).
\end{equation}
This form is based on the (principal) null vector $\that + \rhat$, on
which we perform two orthogonal rotations by the angles $\alpha$ and
$\beta$. In this way, the vanishing norm of the vector is built into
its parametrisation, removing a degree of freedom from the calculation
from the outset. In addition, we obtain a simple picture of the
momentum vector in terms of {\em rotors} (see e.g. \cite{DGL98-grav}).
This form of $p$ yields the following first derivatives of the
coordinates $x^\mu$:
\begin{equation}\label{bl_geod1}
  \tdot  =  \frac{\Phi}{\rho\sDelta}\left(a\sin\theta\sin\beta\sDelta+r^2+a^2\right)\qquad
  \rdot  =  \frac{\Phi\sDelta}{\rho}\,\cos\alpha\cos\beta\qquad
  \thetadot  =  -\frac{\Phi}{\rho}\sin\alpha\cos\beta\qquad
  \phidot  =  \frac{\Phi}{\rho}\left(\frac{a}{\sDelta}+\frac{\sin\beta}{\sin\theta}\right)
\end{equation}
Substituting $p$ into the geodesic equation results in the following
expressions for $\Phidot$, $\alphdot$ and $\betadot$:
\begin{eqnarray}\label{bl_geod2}
  \dot{\Phi} & = &
  \frac{-\Phi^2\cos\beta}{2\rho^3}\Bigg[a^2\sin2\theta\,\sin\alpha + \left(-4ra\sin\theta\,\sin\beta +
  \frac{\rho^2(r^2-a^2)}{r\sDelta} + \sDelta\left(\rho^2-2r^2\right)\right)\cos\alpha\Bigg]\\
  \dot{\alpha} & = &
  \frac{\Phi}{\rho^3\cos\beta}\Bigg[\sin\alpha\bigg(\frac{1}{2r\sDelta}\left(\Delta(\rho^2-4r^2)+\rho^2(r^2-a^2)\right)-2ra\sin\theta\sin\beta\bigg)\nn\\
  && \phantom{\frac{\Phi}{\rho^3\cos\beta}[]}\quad - \cos\theta\cos\alpha\sin\beta\bigg(2a\sDelta+\frac{\sin\beta}{\sin\theta}\left(2(r^2+a^2)-\rho^2\right)\bigg)\Bigg]\\
  \dot{\beta} & = &
  \frac{\Phi}{\rho^3}\Bigg[\cos\theta\sin\alpha\left(2a\sDelta +
  \frac{\left(r^2+a^2\right) + a^2\sin^2\theta}{\sin\theta}\sin\beta\right)\nn\\
  &&\phantom{\frac{\Phi}{\rho^3}[]}\quad-\cos\alpha\sin\beta\left(\frac{r\left(\Delta+a^2\sin^2\theta\right)-m\left(r^2-a^2\cos^2\theta\right)
}{\sDelta}\right) + 2ra\sin\theta\sin\beta\Bigg]
\end{eqnarray}
These equations complete a first order set of 7 equations for the
coordinates $\{t,r,\theta,\phi\}$ and the parameters $\Phi$, $\alpha$
and $\beta$, which is sufficient to determine photon's trajectory and
is numerically very stable. A further advantage of this approach is
that we have not yet made any use of the three conserved Killing
quantities; the energy $E$, angular momentum $J$ and Carter constant
$Q$. In terms of spacetime position and our three photon momentum
parameters, they are given by
\begin{equation}\label{bl_killing}
  E  =  \frac{\Phi}{\rho}(\sDelta + a\sin\theta\sin\beta)\qquad
  J  =  \frac{\Phi}{\rho}\left(\sDelta a\sin^2\theta
    + \sin\theta\sin\beta\left(r^2+a^2\right)\right)\qquad
  Q  =  \Phi^2\rho^2\left(1 - \cos^2\alpha\cos^2\beta\right).\label{bl_carter}
\end{equation}
By evaluating these quantities explicitly at various points along a photon
trajectory we can test that the integration is proceeding correctly.

We apply the same basic idea to calculate the circular, timelike
trajectories in the equatorial plane which model the flow of accreting
matter, except now we can use a single parameter $U$, as follows:
\begin{equation}\label{vpart}
  v = \cosh U \that + \sinh U \phihat
\end{equation}
where the geodesic equations are satisfied by
\begin{equation}\label{bltanhU}
  \tanh U = \frac{\sqrt{r}-a}{\sDelta}.  
\end{equation}
More commonly, these trajectories are written as
\begin{equation}\label{usual_blvpart}
  v \propto \ddt + \Omega \, \ddphi, \qquad \Omega = \frac{1}{a + r^{3/2}}
\end{equation}
and the constant of proportionality is obtained by imposing the
condition $v^2 = 1$. The two notations are related via
\begin{equation}\label{equat_U_and_Omega}
  \Omega = \frac{a + \tanh U \sDelta}{r^2+a^2 + \tanh U\: a\sDelta}.
\end{equation}
In order to model the motion of a source orbiting with the disk
matter, but at a general height $h_s = r\sin\theta$ above the
equatorial plane, we can define a trajectory (no longer geodesic) as
follows
\begin{equation}\label{blvpart_out_of_plane}
  v_s \propto \ddt + \Omega \, \ddphi, \qquad \Omega = \frac{1}{a +
  (r\sin\theta)^{3/2}}, 
\end{equation}
effectively lifting a circular geodesic vertically out of the
equatorial plane. The expression for $U$ given by
Equation~(\ref{bltanhU}) generalises to
\begin{equation}\label{U_out_of_plane}
  \tanh U = \frac{r^2 - a(r\sin\theta)^{3/2}}{a\cos^2\theta +
  (r\sin\theta)^{3/2}}\,\frac{\sin\theta}{\sDelta}
\end{equation}
In order to measure the angle at which a photon leaves the source and
hits the disk we must define a co-moving 3-frame attached to $v_s$
($v$ being a special case where $\theta = \pi/2$. Together with
Equation~(\ref{vpart}), the vectors $\rhat$, $\thetahat$ and
$\phihat^\prime$, where
\begin{equation}\label{bl_corot}
  \phihat^\prime = \sinh U \phihat + \cosh U \that,
\end{equation}
form a convenient orthonormal tetrad.
\bsp
\label{lastpage}
\end{document}